# SYDNEY'S SCIENTIFIC BEGINNINGS:
# WILLIAM DAWES' OBSERVATORIES IN CONTEXT

**Richard de Grijs**
*Department of Physics and Astronomy, Macquarie University,
Balaclava Road, Sydney, NSW 2109, Australia*
Email: richard.de-grijs@mq.edu.au

and

**Andrew P. Jacob**
*Sydney Observatory, Museum of Applied Arts and Sciences,
1003 Upper Fort Street, Millers Point, Sydney, NSW 2000, Australia*
Email: Andrew.Jacob@maas.museum

**Abstract:** The voyage of the 'First Fleet' from Britain to the new colony of New South Wales was not only a military enterprise, it also had a distinct scientific purpose. Britain's fifth Astronomer Royal, Nevil Maskelyne, had selected William Dawes, a promising young Marine with a propensity for astronomical observations, as his protégé. Maskelyne convinced the British Board of Longitude to supply Dawes with a suite of state-of-the-art instruments and allow the young Marine to establish an observatory in the new settlement. The Astronomer Royal may have had a dual motivation, one driven by strategic national interests combined with a personal investment linked to the suggested re-appearance of a comet in the southern sky. With the unexpected assistance of the French Lapérouse expedition, between 1788 and 1791 Dawes established not one but two observatories within a kilometre of Sydney's present-day city centre. Motivated by persisting confusion in the literature, we explore the historical record to narrow down the precise location of Dawes' observatory. We conclude that the memorial plaque attached to Sydney Harbour Bridge indicates an incorrect location. Overwhelming contemporary evidence—maps, charts and pictorial representations—implies that Dawes' observatory was located on the northeastern tip of the promontory presently known as The Rocks (formerly Dawes' Point), with any remains having vanished during the construction of the Harbour Bridge.

## 1 ENTANGLED PERSONAL AND NATIONAL INTERESTS

### 1.1 Lieutenant William Dawes, astronomer on the First Fleet

On 13 May 1787, the 'First Fleet'[1] set sail to Botany Bay (New South Wales, New Holland) from Spithead roads off Portsmouth, England. On 24 October 1786, H.M.S. *Sirius*, a sixth-rate[2] former merchantman equipped with 10 guns—four 6-pounder guns and six 18-pounder carronades—and crewed by a complement of 160, had been commissioned as the convoy's flagship. Under the command of Captain Arthur Phillip (1738–1814; see Figure 1[3]), Governor-designate of the new British colony, the *Sirius*, was accompanied by a faster vessel, H.M.A.T. *Supply*—an armed tender equipped with eight guns (four 3-pounders and four 12-pounder carronades) and a complement of 55 men. In addition, the convoy included three store-ships and six convict transports.

Among the convoy's total complement of 16 officers, 24 non-commissioned officers and 160 privates from the British Marine Corps[4] (Laurie, 1988), as well as 40 women (both convicts and wives of convicts, Marines or crew), Second Lieutenant William Dawes (1762–1836; see Figure 2) had caught the interest of Dr. Nevil Maskelyne (1732–1811), Britain's fifth Astronomer Royal (Howse, 1989). Maskelyne proceeded to train the young man at Greenwich Observatory in the practicalities of astronomical observations. Dawes, already a proficient and competent 'astronomical observer' (Mander-Jones, 1966), had volunteered for service with the First Fleet. He is thought to have received his clearly excellent education in astronomy at the Royal Naval Academy at Portsmouth under headmaster William Bayly (1737–1810; Orchiston, 2016: 152–153), formerly Captain Cook's astronomer (Gibson, 2012). Bayly wrote directly to Joseph Banks (1743–1820; Beaglehole, 1963; Gascoigne, 1994), then-President of the Royal Society of London, to recommend Dawes, "citing his knowledge of languages, botany, mineralogy and also astronomy" (Howse, 2004). Separately, Captain

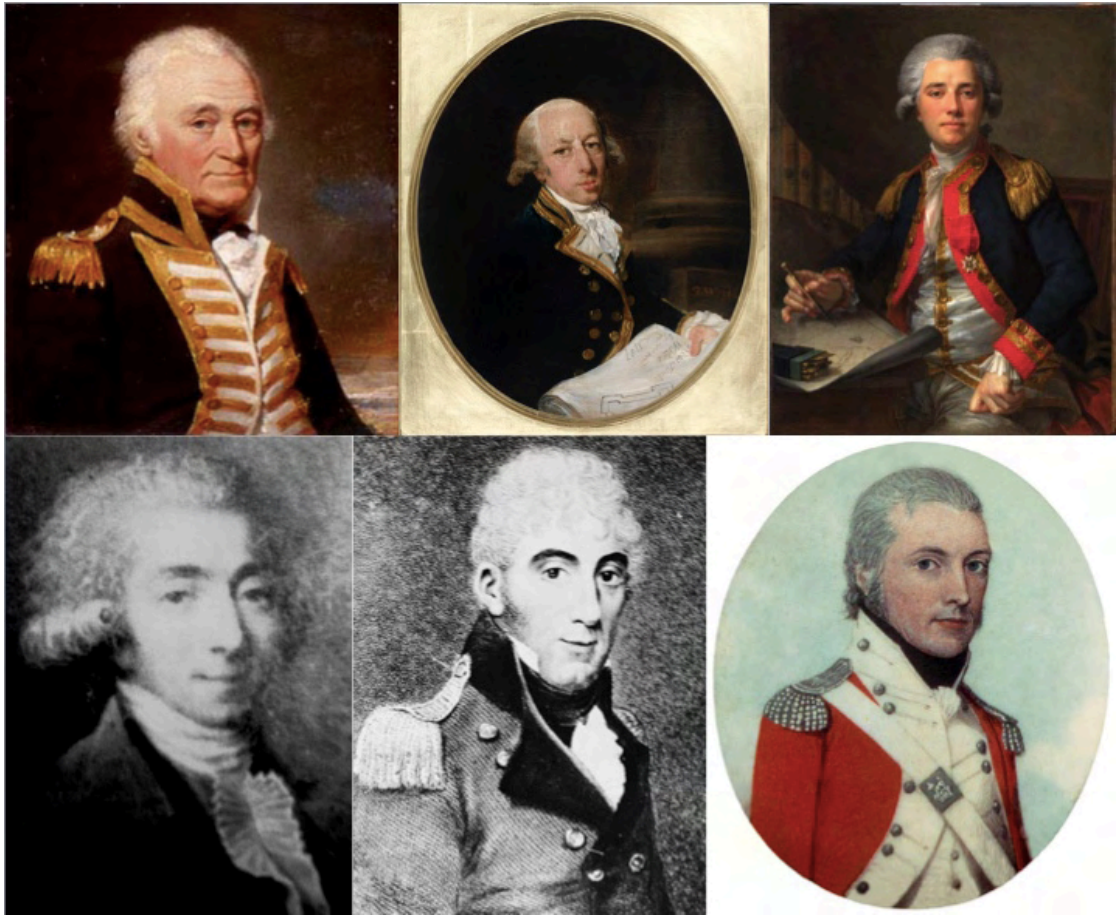

**Figure 1**: Portrait gallery of the main characters described in this article (except for William Dawes), ordered from left to right and from top to bottom by date of birth. Individuals depicted include *(i)* John Hunter (1737–1821; National Library of Australia), *(ii)* Arthur Phillip (1738–1814; Mitchell Library, State Library of New South Wales), *(iii)* Jean-François de Galaup, comte de Lapérouse (1741–1788; Copyright Fine Arts Museums of San Francisco, www.famsf.org, reproduced with permission), *(iv)* Joseph Lepaute Dagelet (1751–1788; courtesy of M$^{me}$ Josiane Dennaud), *(v)* David Collins (1756–1810; National Archives of Australia) and *(vi)* Watkin Tench (1758–1833; Mitchell Library, State Library of New South Wales).

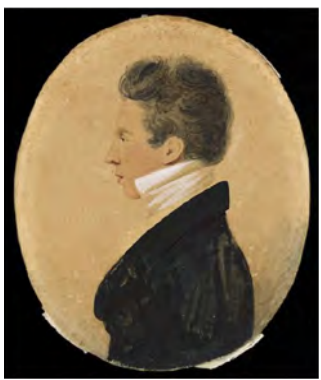

**Figure 2**: Lieutenant William Dawes, 1830s. Miniature oil painting, artist unknown (Tasmanian Museum and Art Gallery, Hobart; ref. AG6048).

William Twiss (1745–1827), Royal Engineer in the British Army, also recommended the young intellectual to Maskelyne (Gillen, 1989). In turn, Maskelyne, who was keen that a capable astronomer be sent to the new colony, highly recommended the young Marine to Banks:

The bearer Mr. Dawes is the person whom I mentioned to you by a letter as desirous to make any observations that might be thought useful at Botany Bay, & whom you approved of in the answer you favored with. He has been a good deal here [at Greenwich Observatory] perfecting himself in the use of astronomical instruments & receiving instructions for discovering and observing the comet of 1532 & 1661 at its first approach to the sun & earth in the summer or autumn of 1788. He is very ready and will do the observing very well. He is desirous of paying his respect, to you, & I desire you will be pleased to allow this letter to be his introduction to you. He will bring you from me a copy of my paper on the return of this comet, which I desire your acceptance of. I saw the comet and got some good observations of it, on the 26th of last month. I have some faint hopes of being able to get a last observation of it tomorrow or some other night soon if very fine before the moon rises. Having parted with [my assistant] Mr. Linley, I at present make all the observations myself, & therefore cannot attend the first meeting of the

Royal Society. (Maskelyne, 1786b)

While Dawes was clearly a bright and intelligent young man, Maskelyne's choice of him as his emissary nevertheless seems odd, given his working-class background. As the nation's pre-eminent astronomer, Maskelyne was not short of well-qualified, university-educated volunteers, so Dawes must indeed have come with the highest praise. Dawes' successful appointment by the Home Secretary, Thomas Townshend, First Viscount Sydney, and the Board of Longitude is even more surprising given the enormous importance afforded to establishing the new colony. Other luminaries, including Banks and Sir Evan Nepean (1752–1822), Permanent Undersecretary of State for the Home Department, also vied for patronage. Indeed, Maskelyne seems to have outwitted his nemesis Banks in securing Dawes' loyalty (e.g., Gibson, 2012).

From our current perspective, it may seem odd that sending an astronomer to the yet-to-be-established colony—even if he was also a Marine, a competent engineer, surveyor and explorer—was considered a high priority. Shortages of labour, materials and food would hamper the colony's early development and keep Governor Phillip occupied, yet Dawes was meanwhile commissioned to establish an astronomical observatory.

The focus of this article is on Dawes' observatory in New South Wales,[5] the developments leading up to its foundation (Section 2), the instruments it was equipped with and the role it played in the social fabric of the new colony (Section 3). While some of this story is reasonably well known, our aim here is to provide a comprehensive account of the history and operations of the first permanent[6] British astronomical observatory in the southern hemisphere. We have uncovered a number of new insights, and we have also established the observatory's site beyond any reasonable doubt based on a careful assessment of the historical record (Section 4). Our historical research has been based on a wide range of records. Among the online documents we consulted were the full set of papers of the Board of Longitude and eyewitness accounts of the First Fleet's voyage. In addition, we explored the online and physical holdings of the State Library and State Archives of New South Wales, the library of the Royal Australian Historical Society (RAHS), all in Sydney, and the National Library of Australia in Canberra.

**1.2 An astronomer on the 'First Fleet'**

In the late eighteenth century, the dominant sea-faring European nations continued to open up commercial shipping routes to ever more distant destinations, yet a reliable means of geographic position determination at sea remained elusive. Determination of one's latitude was relatively straightforward: one simply needed to measure the height of the Sun or those of one or more bright stars at their meridian passage, corrected for seasonal variations, and a latitude measurement would follow naturally.

Determination of one's longitude at sea was significantly more complicated (for a recent review, see de Grijs, 2017; see also Andrewes, 1996). Longitude determination relies on knowing one's local time with respect to that at a reference location (such as the Greenwich meridian). Solving the longitude problem occupied generations of scientist-scholars and navigators. It took until the second half of the eighteenth century before John Harrison (1693–1776) had perfected his maritime timepiece to work sufficiently accurately that he was awarded a significant fraction of the British Longitude Prize of 1714. Harrison's payment was delayed, however, until the British watchmaker Larcum Kendall (1719–1790) had managed to successfully copy Harrison's H4 in 1769. Kendall's copy is now referred to as his No. 1 Marine Timekeeper, 'K1'. James Cook (1728–1779) took K1 on his second and third voyages of discovery to the South Pacific, calling it "our trusty friend the Watch" and "our never failing guide the Watch" (Andrewes, 1996: 226, 252). Indeed, after more than three years at sea, in 1775 it was determined that the timepiece had gained just 13 seconds a day on Cook's second voyage (e.g., Howse and Hutchinson, 1969; Hawkins, 1979).

Chronometers were few and far between, however, and they were expensive to obtain. Alternative means of longitude determination had been developed, most involving angular distance measurements between the lunar limb and the positions of bright stars in the sky, an approach known as the 'lunar distance' method (for a review, see de Grijs, 2020). Starting in 1767, Maskelyne had been instrumental in publishing his *Nautical Almanac*s, which included lengthy tables of lunar distances in time intervals of three hours for Greenwich time. Shipboard lunar distance measurements, usually based on observations with a sextant, would allow accurate determination of the time in Greenwich. This could then be compared with the observer's local time to yield a longitude measurement. The calculations required to carry this process through to completion were highly complicated and tedious, however.

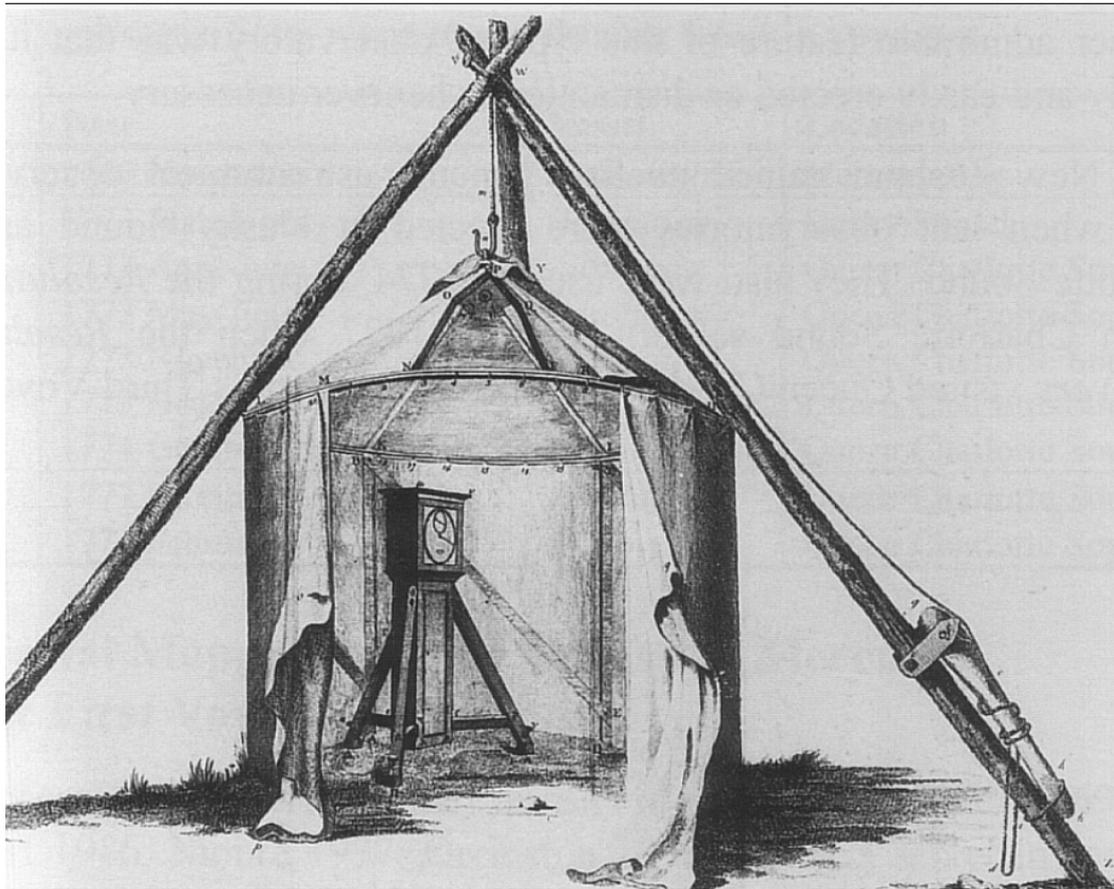

**Figure 3**: Typical tent observatory used by shipboard astronomers in the seventeenth and eighteenth centuries (Wales and Bayly, 1777: Plate II).

For these and other reasons, voyages of discovery usually accommodated one or more competent astronomers among their crews. These itinerant astronomers were often equipped with portable 'tent' observatories (see Figure 3). In the seventeenth and eighteenth centuries, increasingly accurate astronomical observations from across the globe were urgently needed to improve shipping practices. In fact, Maskelyne complained in the 1788 *Nautical Almanac* that …

> It is indeed to be lamented that Persons, who visit distant countries, are not more diligent to multiply Observations of this kind; for want of which, the Observations made by Astronomers in established Observatories lose half their Use and the improvement of Geography is retarded. (Maskelyne, 1787)

Maskelyne thus convinced Lord Sydney and the Board of Longitude to support an astronomical component for the First Fleet, including the provision of a range of instruments and books on loan for the colony's new observatory. These would ostensibly be used for determination of the settlement's precise position, to accurately determine the colony's local time, to provide an accurate time reference for the chronometers of visiting ships and to assist with surveying duties. However, from Maskelyne's personal letters we also get the distinct impression that he was personally invested in a positive outcome of the astronomical enterprise. Maskelyne had convinced the Board of Longitude to send Dawes to New South Wales so as to test the Astronomer Royal's prediction of the return of the comet of 1532 and 1661 to the southern sky. The comet's earlier recorded appearances had led Edmond Halley, Britain's second Astronomer Royal, to predict that it would reappear in 1790. However, in a paper read to the Royal Society of London on 29 June 1786, Maskelyne suggested that the comet might reappear as early as 1788 or 1789, with a perihelion (its closest approach to the Sun) potentially as early as 1 January 1789:

> It will approach us from the southern parts of its orbit, and therefore will first appear with considerable south latitude and south declination; so that persons residing nearer to the equator than we do, or in south latitude, will have an opportunity of discovering it before us. It is to be wished that it first be seen by some astronomer in such a situation, and furnished with proper instruments for settling its place in the heavens, the earliest good observations being most valuable for determining its elliptic orbit … The Cape of Good Hope would be an excellent situation for this purpose. (Maskelyne, 1786a)

If Maskelyne's prediction were correct, this would be the second comet to be predicted correctly. The return of Halley's comet had been confirmed in 1759 following the astronomer's published prediction of 1705 (Halley, 1705), and so Maskelyne was particularly keen to make his mark alongside Halley. Therefore, he recommended that Dawes be appointed as the official astronomer on board the *Sirius*, with an expanded brief to 'recover' Maskelyne's comet and establish an observatory in the new colony. He persuaded the Board of Longitude to also provide Dawes with a suitable complement of instrumentation to allow him to do his job successfully. Astronomy—whether a personal pursuit or of national strategic importance—thus became a foundational component of the new colony in New South Wales. As we will see shortly, despite concerted efforts, Dawes did not find the comet. Astronomers in Europe did not find it either, thus suggesting that Maskelyne's predictions may have been wrong.

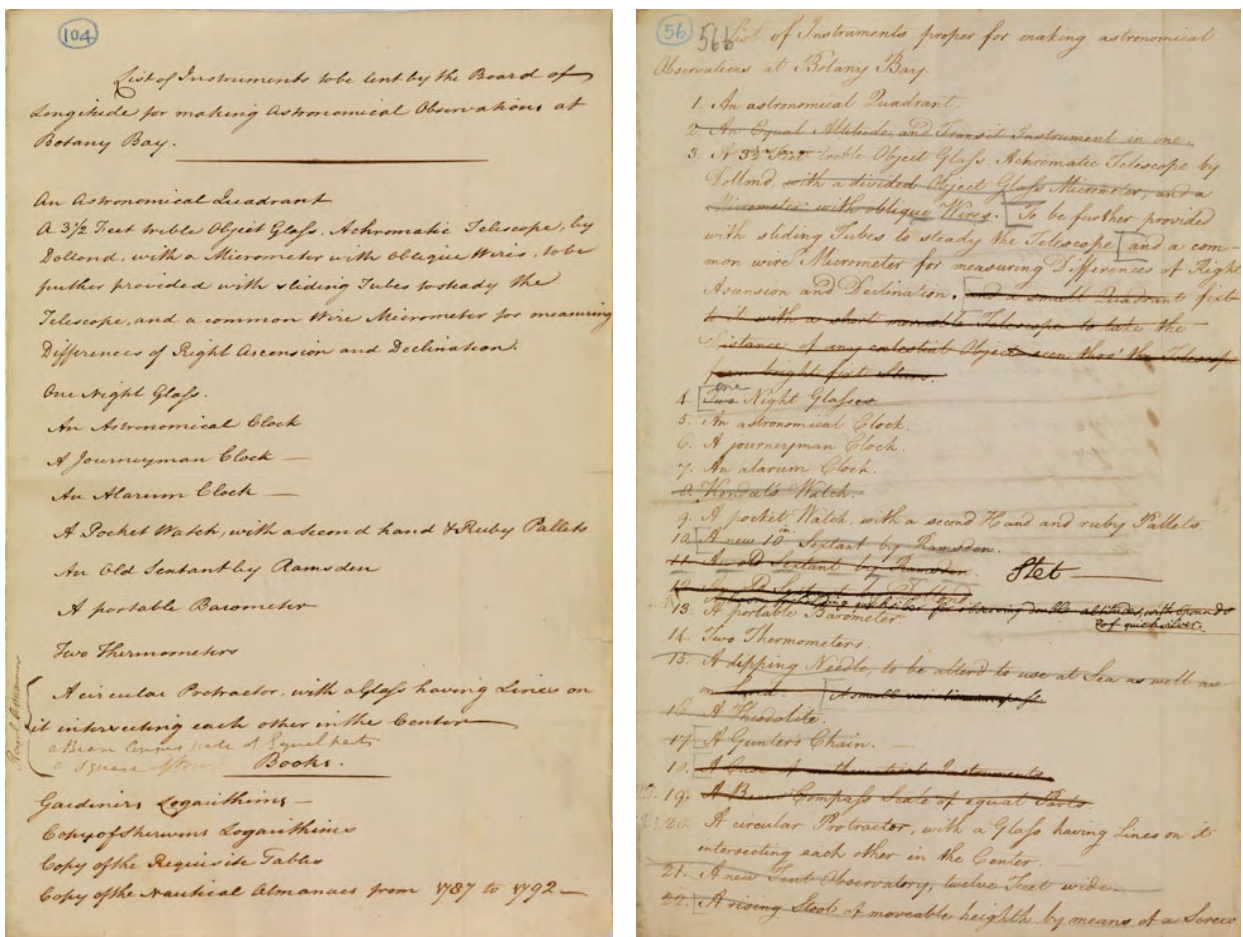

**Figure 4**: (*left*) Dawes' 'List of Instruments proper for making astronomical Observations at Botany Bay'. (*right*) 'List of Instruments to be lent by the Board of Longitude for making Astronomical Observations at Botany Bay' (National Museum of Australia).

**1.3 Instrumentation for the new observatory in New South Wales**

Maskelyne convinced the Board of Longitude to lend the expedition a set of astronomical and meteorological books and equipment, selected by the Astronomer Royal to closely resemble the instrumentation carried on Cook's already legendary voyages of the late 1760s and 1770s (Howse and Hutchinson, 1969; Howse, 1979). Dawes' initial request, submitted in early November 1786 and contained in the 'List of Instruments proper for making astronomical Observations at Botany Bay' (Figure 4, left; Dawes, 1786), included a large number of instruments. However, the Royal Society was unable to source all instruments requested prior to the First Fleet's departure, so that a significant number were crossed out. Eventually, the expedition was provided with a set of instruments and books, under the care of Captain Phillip, contained in the much shorter 'List of instruments to be lent by the Board of Longitude for making astronomical Observations at Botany Bay' (Figure 4, right):

- A [12 inch] Astronomical Quadrant [by John Bird, for longitude determination on land]
- A 3½ Feet [~1 m] treble Object Glass Achromatic Telescope by Dollond with a micrometer with oblique wires, to be further provided with sliding tubes to steady the Telescope, and a common wire Micrometer for measuring Differences of Right Ascension and Declination [presumably a compact, portable 'comet sweeper', owned by Maskelyne]
- One Night Glass
- An Astronomical Clock [by John Shelton]
- A Journeyman Clock [probably also by Shelton]
- An Alarm Clock [maker unknown]
- A Pocket Watch, with a second hand & Ruby Pallets [possibly by Ellicott, No. 4659]
- An old [Hadley's] sextant by Ramsden
- A portable Barometer [by Burton]
- Two Thermometers
- A circular Protractor with a Glass having lines on it intersecting each other in the center [*sic*]

<u>*Books*</u>

- [William] *Gardiner's Logarithms* [*Tables of Logarithms*, 1742]
- Copy of [Henry] *Sherwin's Logarithms* [*Sherwin's Mathematical Tables ... Table of Logarithms of Numbers from 1 to 101*]
- Copy of [Maskelyne's] *Requisite Tables*
- Copy [*sic*] of the *Nautical Almanac*s from 1787 to 1792

Although Dawes was initially comfortable with Phillip's guardianship of the instruments, he expressed some concern to Maskelyne that his need of access to the *Nautical Almanac* might interfere with the Captain's requirements. In response, Maskelyne provided him with a second set of books. Dawes may have taken some of his own equipment on the voyage as well, and he was given some additional instruments by Colonel Robert Jacob Gordon (1743–1795), commander of the Dutch troops in Cape Town, for use upon arrival at Botany Bay (James, 2012). There are no surviving records of any such equipment, however.

Among the instruments taken to the new colony (for a detailed discussion, see Laurie, 1988: 470–471), the clocks merit further discussion. The Board of Longitude had provided the expedition with both an astronomical pendulum clock and a journeyman or assistant clock. The latter was equipped with a gong that sounded every minute, while the seconds could be counted based on its ticking sound (James, 2012). The astronomical clock is thought to have been made by the specialist English regulator maker John Shelton (1712–1777). It was purchased in 1768 for £40 sterling and served as main reference of either sidereal or local time on Cook's second and third voyages to the South Pacific (see also Orchiston, 2016). This clock, most likely Royal Society Regulator No. 35 and now a treasured item in the British National Maritime Museum's collection in Greenwich (Hawkins, 1979), showed hours, minutes and seconds to high accuracy.

On board the *Sirius*, Captain Phillip was also entrusted with the special care of Kendall's K1 chronometer. Prior to the First Fleet's departure from England, special instructions were provided, specifying that K1 had to be wound every day at noon. Despite the importance of retaining Greenwich time to high accuracy during the entire voyage, shortly after the convoy's departure from Cape Town, when Phillip and Dawes transferred from the *Sirius* to the *Supply* to expedite their arrival in Botany Bay, K1 was left to wind down. Dawes was forced to regain the Greenwich time reference based on lunar distance measurements (James, 2012):

> This separation, the first that had occurred, did not take place until the 25th [November 1787], on which day Captain Phillip went on board the *Supply* taking with him, from the *Sirius*, Lieutenants [Philip Gidley] King and Dawes, with the [K1] time-keeper. (Collins, 1798: xxxv)

Once Dawes had established an observatory in New South Wales, he would have routinely checked and recalibrated K1's time measurement against that shown by his Shelton clock (e.g., Hawkins, 1979).

## 2 AN OBSERVATORY AT PORT JACKSON

### 2.1 An encounter with the French

H.M.A.T. *Supply*, carrying Captain Phillip and Dawes, arrived in Botany Bay[7] on 18 January 1788, the following day followed by the fastest convict transports—*Alexander*, *Friendship* and *Scarborough*. The remainder of the First Fleet, including H.M.S. *Sirius*, arrived on the 20th. The settlers soon realised, however, that Botany Bay was not their hoped-for site to establish a permanent settlement.

An exploration party led by Phillip, also including John Hunter (1737–1821), Second Captain of the *Sirius*, sailed from Botany Bay on 21 January and discovered that Port Jackson[8], some 12 km further north, provided adequate shelter, deep water for the ships' anchorage, fresh water and ideal conditions for agriculture (e.g., Parker, 2009). In Phillip's words, Port Jackson represented "the finest harbour in the world, in which a thousand sail of the line may ride in the most perfect security …" (Phillip, 1790: 55). Surgeon Arthur Bowes-Smyth—known simply as 'Bowes' in the colony—said,

> The finest terras's [*sic*], lawns and grottos, with distinct plantations of the tallest and most stately trees I ever saw in any nobleman's ground in England, cannot excel in beauty those wh. Nature now presented to our view. The singing of the various birds among the trees, and the flight of the numerous parraquets, lorrequets, cockatoos, and macaws [note that macaw habitats do not exist in Australia], made all around appear like an enchantment; the stupendous rocks from the summit of the hills and down to the very water's edge hang'g [*sic*] over in a most awful way from above, and form'g [*sic*] the most commodious quays by the water, beggard [*sic*] all description. (Fidlon and Ryan, 1979; Hill, 2015: 107).

Following the party's return to Botany Bay on 23 January 1788, the convoy prepared to depart for Port Jackson. However, "to the infinite surprise of everybody, we saw two large ships in the offing" (Bowes-Smyth, 1790: 87). Unbeknownst to the British, a French scientific expedition led by Jean-François de Galaup, comte de Lapérouse (1741–1788; Marchant, 1967), had arrived just outside Botany Bay. The French were equally surprised, as we learn from Lapérouse's journal:

> We spent the whole of the 24th [January 1788] plying in sight of Botany Bay, but that day we had a spectacle; this was the British Squadron at anchor with their pennants and ensigns which we could plainly distinguish. (Dunmore, 1995)

The French expedition counted 10 scientists among its complement of 114, including the astronomer Joseph Lepaute Dagelet (1751–1788), whom we will meet again shortly. In fact, the arrival of the French frigates, *Astrolabe* and *Boussole* ('Compass'), added urgency to Dawes' and Captain Phillips' orders to establish an observatory in the new colony. Lapérouse's Royal instructions of 15 February 1785 included such orders explicitly:

> Immediately upon arriving in a harbour, he will select an appropriate site on which to erect the tents and the observatory, and will set up a guard. … Separately from observations relating to the determination of latitudes and longitudes, for which every known and practicable method will be used, and those needed to assess the declination and inclination of the dipping needle [a freely suspended magnetic needle used to determine the local magnetic inclination], he will ensure that any celestial phenomenon which may be visible be observed; and on every occasion he will give the astronomers all the help and facilities necessary for the success of their work. (Dunmore, 1994: cxlii–cxliii)

Despite gale conditions, the First Fleet weighed anchor on 25 January 1788, *en route* to Port Jackson. The site the exploration party had selected for settlement, some 10 km into the natural harbour, was known as War-ran (Warrane) by the Indigenous Gadigal clan. Today, the inlet adjacent to Sydney's central business district is known as Sydney Cove. The cove featured sheltered deep water close to the shore and a fresh-water tributary (now known as the Tank Stream), where "ships can anchor so close to the shore that, at small expense, quays may be built at which the largest ships can unload" (Phillip, 1790: 55). Indeed, today the Overseas Passenger Terminal at Sydney's Circular Quay routinely hosts some of the world's largest cruise ships.

Dawes was keen to start construction of the settlement's observatory. However, from his letters to Maskelyne we learn that his duties as Marine officer on the *Sirius* interfered (Dawes, 1788a). His application for a shore assignment was initially refused, although he was promised that he would be considered for the first land-based vacancy to arise (Mander-Jones, 1966).

Meanwhile, the French had built a stockade, an observatory and a garden for fresh produce on the headland now known as La Perouse, on Botany Bay's northern shore.[9] On behalf of Governor Phillip, Dawes and Lieutenant Philip Gidley King (1758–1808) paid their French counterparts a visit on 2 February 1788, offering them any assistance they might need. They left Sydney Cove at 02:00 that morning for an eight-hour sailing (and rowing) voyage (Fidlon and Ryan, 1980: 37–40; Clarke, 2015: 33), battling strong southerly winds on their way to Botany Bay. We learn from King's private journal that the temporary French settlement was already set up for astronomical observations to be taken:

> After dinner I attended ye Commodore & other Officers onshore where I found him [Lapérouse] quite established, having thrown round his Tents a Stoccade, guarded by two small guns in which he is setting up two Long boats which he had in frames, an observatory tent was also fixed here, in which were an Astronomical Quadrant, Clocks &c under the Management of Monsieur Dagelet Astronomer, & one of ye Académie des Sciences at Paris.[10] (King, 1788)

The quadrant, one of five made by Claude Langlois (ca. 1700–1756; the most highly regarded scientific instrument maker in France between about 1730 and 1756), was the most important of Dagelet's instruments. It was on loan from Paris Observatory. In addition, the French observatory at Botany Bay was equipped with a meridian telescope of a design developed by Alexis-Marie Abbé de Rochon (1741–1817), an invariable pendulum for gravity measurements of the type Charles-Marie de La Condamine (1701–1774) had used in the Americas—which was 'invariable' since it had a fixed pendulum length—three astronomical pendulum clocks made by the celebrated clockmaker Jean-André Lepaute (1720–1789), Borda reflecting circles (see, e.g., de Grijs, 2020) and English-manufactured sextants (see, e.g., Barko, 2007). However, the observing conditions at Botany Bay were far from ideal. On 6 February 1788, Dagelet wrote to Marie Jean Antoine Nicolas de Caritat (1743–1794), Marquis de Condorcet and permanent secretary of the Académie Royale des Sciences, "I am blinded by the bites of flies which weigh me down in my wretched observatory" (Morrison and Barko, 2009: 28, note 14).

Dawes and Dagelet, the more senior scholar, thus met in person for the first—and only—time on 2 February 1788. As men of science, their conversation naturally turned to Dawes' assignment to establish an observatory in the new settlement and to Cook's nautical and astronomical observations. The French unreservedly praised the accuracy of Cook's measurements, with Lapérouse adding magnanimously, "Mr. Cook did so much that he left me nothing to do but to admire his work" (Barton, 1889).

The French and English contingents maintained cordial relations during Lapérouse's six-week sojourn on the shores of Botany Bay. Although Dagelet never visited Dawes at his Port Jackson base, Lieutenant Charles-Marie Fantin de Boutin (born ca. 1760) visited Sydney Cove and reported back to Dagelet. De Boutin had inspected the foundations of Dawes' observatory, most likely around 21 February, on a visit with Father Jean-André Mongez (1750–1788), the French expedition's chaplain (for the relevant journal entries, see Protos, 1988).

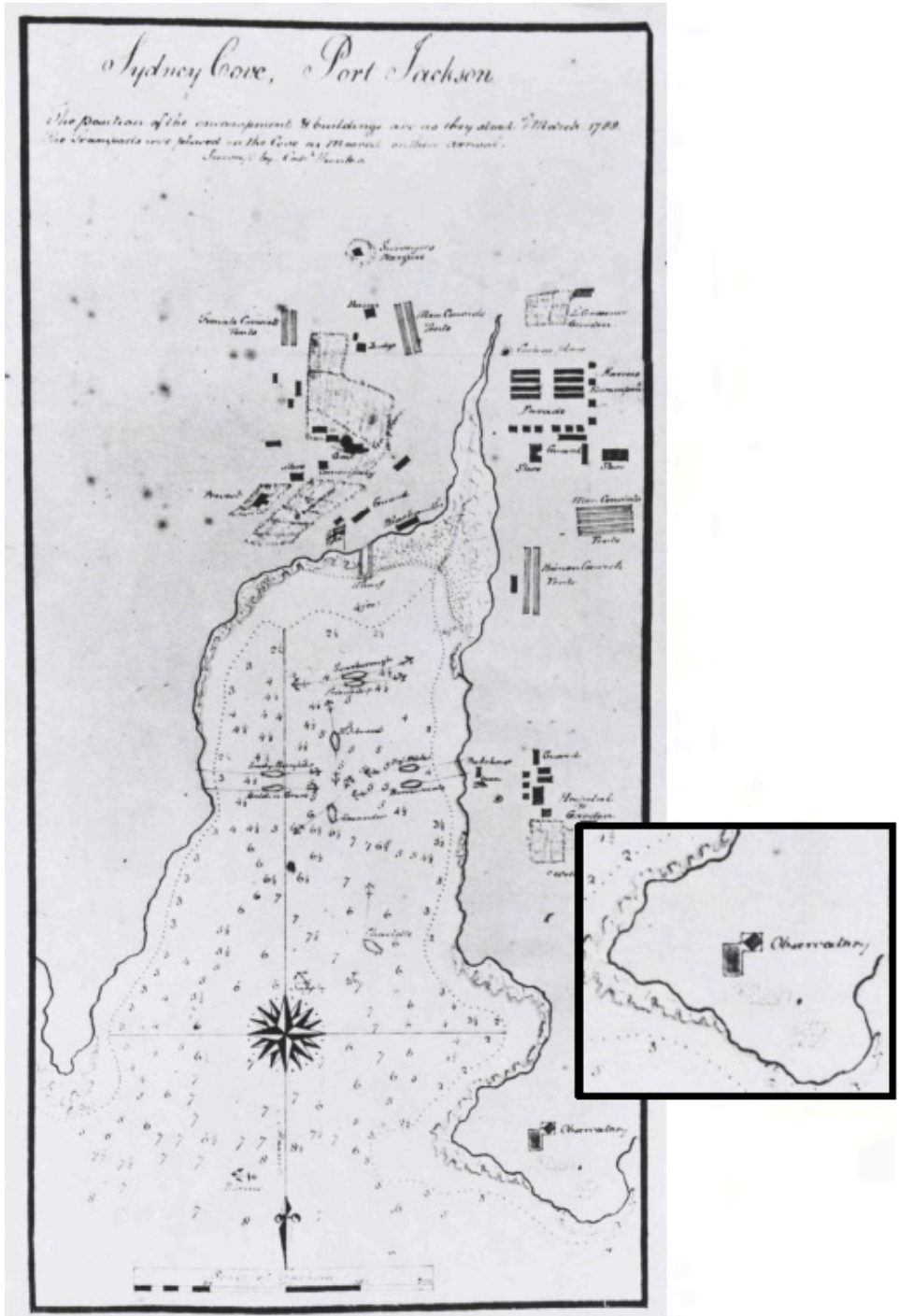

**Figure 5**: "Sydney Cove, Port Jackson, surveyed by Captain Hunter, 1 March 1788." For a high-resolution image, see https://nla.gov.au/nla.obj-230724226/ (National Library of Australia).

**2.2 An observatory for the colony**

On 3 March 1788, Dagelet sent Dawes a letter[11] that offered advice on how to set up his observatory, which was partly based on de Boutin's description of the most likely site. Dagelet's letter included suggestions about the observatory's layout, the placement of Dawes' instruments and ideas for future instruments—including, ambitiously, "a sector [transit circle] with a radius of 6 to 8 feet" (1.8–2.4 m). He also provided ideas as to the types of observations Dawes might pursue, particularly encouraging him to attempt "the measurement of a degree of the meridian in this hemisphere" (Barko, 2007). Despite their brief acquaintance, both men clearly got along well (e.g., Barko, 2007). In a letter from Dagelet to his mentor, the eminent Parisian astronomer Joseph Jérôme Lefrançois de Lalande (1732–1807), which he asked Dawes to send via Maskelyne, Dagelet sounded hopeful, advising that

> … at Botany Bay he had come across an English astronomer furnished with instruments who was preparing to carry out numerous observations and we may have the satisfaction of correspondence with our Antipodes. (Barko, 2007: 21)

It took until March 1788 before Dawes would be assigned to a substantial shore-based role in Governor Phillip's quest to secure the territory for the British Crown. He was appointed[12] as the settlement's engineer, artillery officer[13] and surveyor, roles that allowed him shore access, although he remained part of the *Sirius*' complement of Marines until early July of that year. Nevertheless, already in February 1788 do we learn of Dawes' efforts to commence construction of an observatory:

> Among the buildings that were undertaken shortly after our arrival must be mentioned an observatory, which was marked out on the western point of the cove, to which the astronomical instruments, which had been sent out by the Board of Longitude for the purpose of observing the comet which was expected to be seen about the end of this year. The construction of this building was placed under the direction of Lieut. Dawes, of the Marines, who, having made this branch of science his peculiar study, was appointed by the Board of Longitude to make astronomical observations in this country. (Collins, 1798: 15)

Hunter's first map of Sydney Cove (see Figure 5), dated 1 March 1788, already shows the observatory's site and an anticipated outline of a two-building structure. The observatory building appears large compared with other contemporary buildings. In comparison with modern maps, the location of the Governor's residence is shown close to its actual position, whereas the topography of Sydney Cove maps well onto the modern shoreline (see below). Meanwhile, land-based astronomical observations were obtained from a portable tent observatory similar to that carried by Cook and his astronomer, Charles Green (1734–1771), on the H.M.B. *Endeavour*'s round-the-world voyage of 1768–1771 (e.g., Haynes et al., 1996: 31–32). Assisted by four marines and a few convicts, construction of a wooden observatory had well and truly commenced by 20 April 1788:

> I have notwithstanding with the assistance of four marines of my own party and three or four convicts when the Governor has allowed me, cleared a point of land of trees, and am now getting on as fast as possible with an observatory which I hope will be completed and the instruments in it by July sometime. This has not however been done without a good deal of my own and my servants' bodily labour which shall when necessary be cheerfully employed in the same cause. The situation struck me at first sight to be so eligible and all the necessary materials so conveniently at hand, that I did not hesitate a moment to determine on setting about a permanent one at once <u>a sketch will say more than many words</u>. (our emphasis: see Figure 6; Dawes, 1788a)

Dawes had been given permission to build the structure on the thickly wooded, rocky promontory less than a kilometre from the settlement on the western side of Sydney Cove. It was known to the Gadigal people as Tar-ra or Tjara (Dara).[14] The site of Dawes' Observatory was separated from the main settlement by a track along the waterfront. At Dawes' request, Hunter "was pleased to honour this Point [Tar-ra] by calling it Point Maskelyne" (Dawes, 1788e). However, the area was colloquially referred to as Dawes' Point, which remains in use today. (We will discuss the observatory's location in more detail in Section 4.) The Gadigal people considered the headland on the western side of Sydney Cove a safe and welcoming location to share friendship and knowledge (Moran and McAllister, 2020).

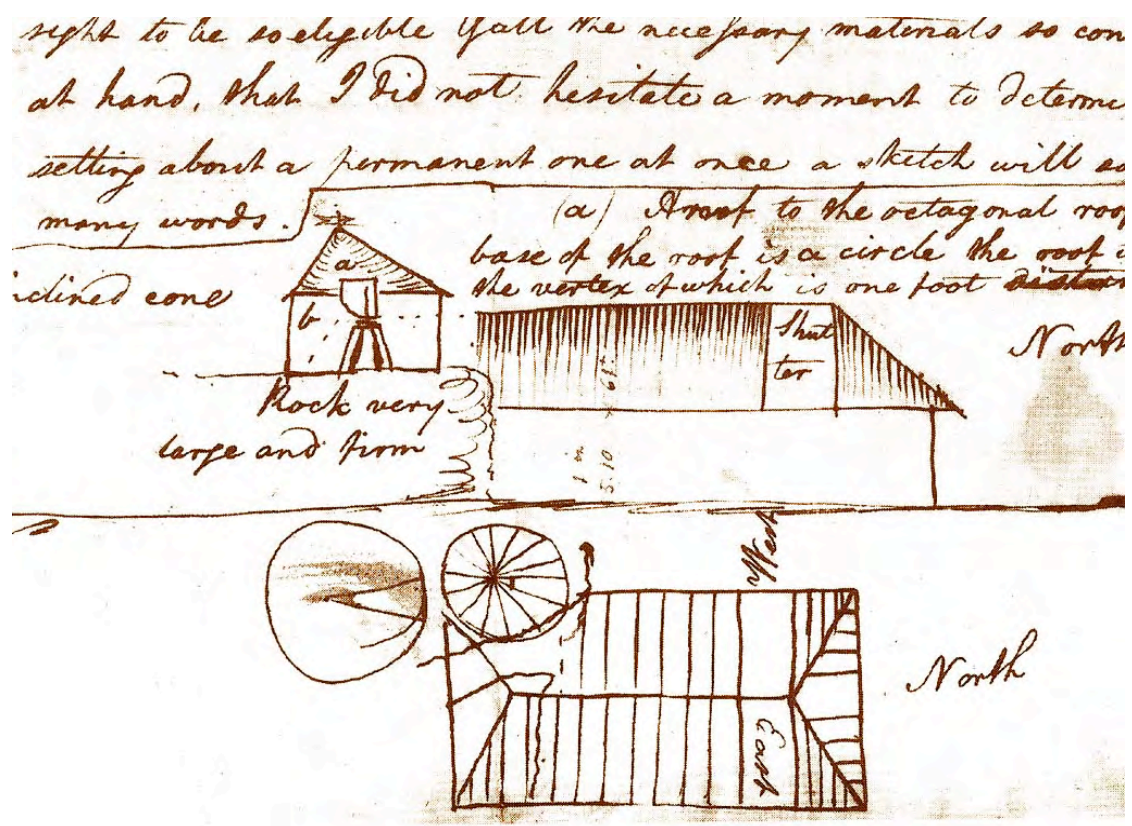

**Figure 6**: Notebook sketches by William Dawes of his cottage and observatory on Point Maskelyne (Cambridge University Library; Board of Longitude Papers, RGO 14/48 f. 281rv).

Construction of the observatory proceeded well those first few months, despite a chronic shortage of materials. In fact, Phillip confided in Dawes that if he had known the difficulties they would face in sourcing building materials, he would have opted to bring an expensive, pre-fabricated observatory building along on the voyage from England (Dawes, 1788b). Dawes managed to motivate his construction crew by providing his Marines with shoes and rum and water "now & then" during the worst of the summer heat. By July 1788, fellow Marine Watkin Tench (1758–1833) reported that the observatory was

> … nearly completed, and when fitted up with the telescopes and other astronomical instruments sent out by the Board of Longitude, will afford a desirable retreat from the listlessness of a camp evening at Port Jackson. (Tench, 1789)

Despite the delay incurred by a lack of suitable materials and qualified tradesmen in the colony, Dawes moved into the new observatory's living quarters in the first week of July. Governor Phillip provided an update on the observatory's construction in his letter to Lord Sydney of 9 July 1788:

> On the point of land which forms the west side of the cove an Observatory is building, under the direction of Lieutenant Dawes, who is charged by the Board of Longitude with observing the expected comet. (Phillip, 1788)

In his letter to Maskelyne of 10 July 1788, Dawes (1788b) is clearly excited about the progress he has made. He told Maskelyne that he was considering extending his time in the colony by signing on for another three years. However, having said so, just two days later he placed a number of conditions on his request for an extension. Most importantly, he declared that he would need to be given a more permanent appointment at the observatory to supplement his Marine income:

> With respect to remaining in this Country I have to observe that I shall, during the three Years for which I offer'd my Services, pay all possible attention to the observatory; & that, from a wish to make every return in my power, for the very polite and friendly manner in which you have been pleased to treat me; but, as staying here without any certainty of some addition to my income would be doing myself and Family a material injury; I propose, unless I receive some assurance of such addition, to return at the end of that term. It is not that I wish an allowance from the B.ᵈ of Longitude in particular, but if any appointment can be thought of in any department, which may tie me to this Country, I shall with the greatest pleasure attend to the observatory to the utmost of my power without thinking of any stipulation with the board of Longitude in particular. Nevertheless, as I have had the happiness of being in some measure the means of establishing a permanent observatory in this country, I should, if a fix'd observer be thought necessary be equally happy to fill that place. If not, I will (while I can with common justice remain in this country) attend to the observatory as strictly as if I were paid a salary for that business, … (Dawes, 1788c)

Although Dawes did not state it directly at this time, later that year he expressed his conviction that he had given the observatory as much as he could: "no application shall be wanting in me" (Dawes, 1788e). He hoped that a properly qualified successor could be found by the time his six years of service were up. Clearly, he had meanwhile decided to sign on for a second three-year term. In fact, the historical record offers us a useful insight into his thinking: in a letter to his father he admitted that the "principal reason" to sign on for a second term was

… the uncertainty of any other person being appointed to succeed him in the astronomical way, and the great loss, in his opinion, it would be to the science were the opportunity to be missed of making proper use of those valuable instruments. (Bladen and Britton, 1893: 423)

Given the observatory's almost-operational state by mid-July 1788, Dawes arranged for his instruments to be transferred to shore, unpacked and ready for use by the first week of August. He was keen to start his search for Maskelyne's comet (Dawes, 1788a). Routine scientific operations commenced on 14 September 1788.

**3 COMPLETION OF THE OBSERVATORY AND ROUTINE OPERATIONS**

**3.1 Building specifications**

Dawes' timber and canvas observatory was, as expected for a building constructed during the colony's first year of settlement, small and far from impressive. Nevertheless, its completion represented a milestone for the young colony, given that it was the first building dedicated to astronomy and meteorology in New South Wales.

Dawes' observatory consisted of two buildings made of timber, one of which was octagonal, nine feet (2.7 m) in diameter, with a white conical canvas roof and built on a "very large and firm" rock (James, 2012):

I have got an exceeding good stone cut into the form of a frustum of an octagonal pyramid whose base is 2 ft 5 in [~74 cm] and top 1 ft 5 in [~43 cm] this to be placed on the centre of the octagonal room & the quadrant on the top of it. The roof is to turn round on three or more rollers. *e*, A stair case of communication between the upper & lower room. *f*, the proposed place for the astron[omical] clock. The roof of the lower room is to be so constructed and to make good against the side of the upper room and the ridge of it is to be several inches below the top of the quadrant. (Dawes, 1788a)

The revolving canvas roof on the octagonal building included a retractable shutter. The roof's apex was located approximately a foot (~30 cm) from the vertical position, so that astronomical observations could also be made at zenith (James, 2012). The canvas roof was nailed to vertical wooden poles that rested on cannon balls in a gouged wooden track to allow for rotation (Haynes et al., 1996: 31–32). It is possible that the tent observatory used during the voyage and upon their arrival in Port Jackson was repurposed to provide the canvas roofing material (Kerr, 1986).

Dawes selected a large outcrop of either bedrock or a large 'floater' as the solid foundation for his quadrant room. The quadrant required a foundation that was stable against vibrations, "on a good stone ... an octagonal pyramid", so as to work effectively. This thus determined the final site selection for the observatory (Saunders, 1990) at the northeastern end of the Tar-ra headland, northwest of what is now Campbells Cove.

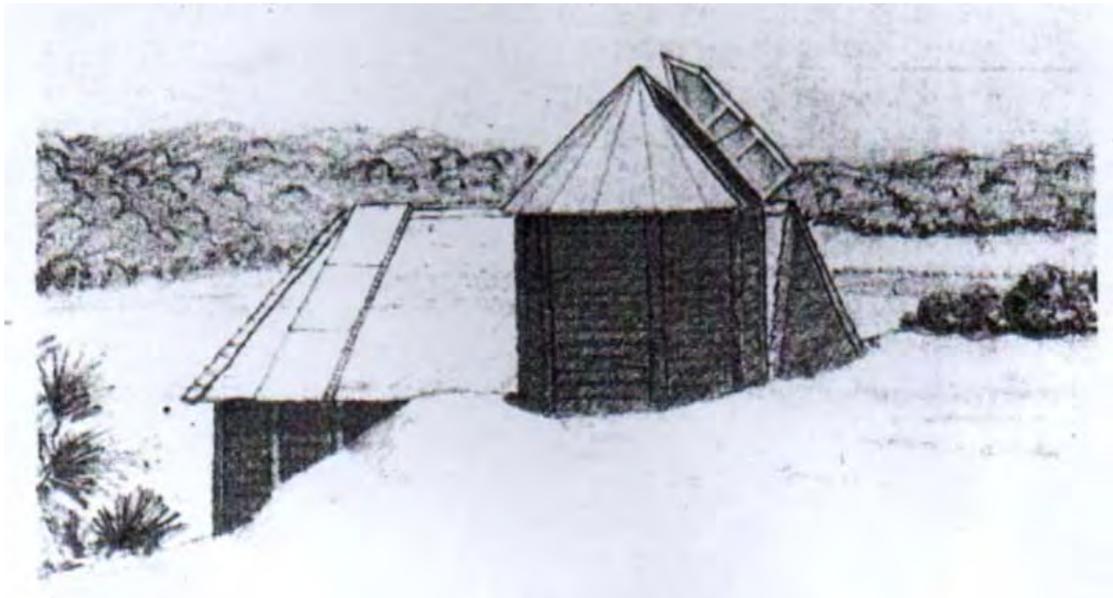

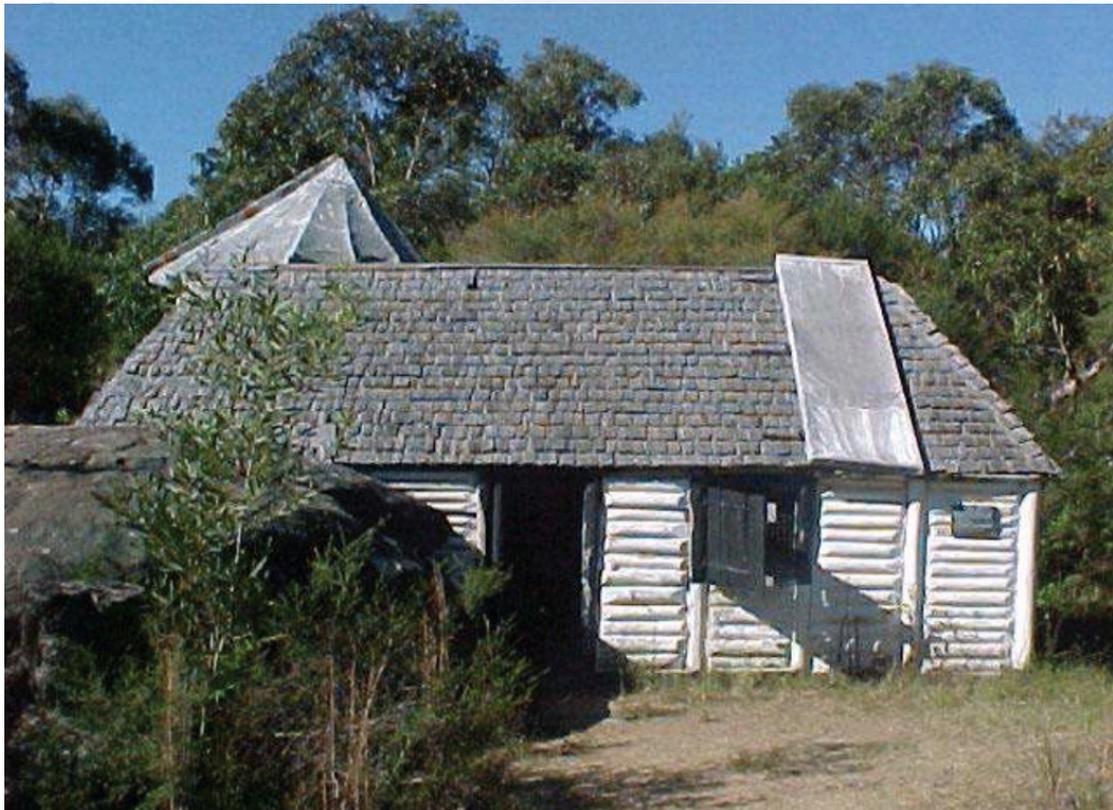

**Figure 7**: (*top*) Sketch of Dawes' observatory at Sydney Cove (Credit: Rod Bashford; McAfee, 1981). (*bottom*) Reconstruction of Dawes' observatory at Old Sydney Town.

A second, rectangular building measuring 12 by 16 feet (3.7×4.9 m$^2$) resembled a lean-to shed and was constructed against that rock (Dawes, 1788a; see also Figure 7, top). It served as Dawes' study and living quarters. A staircase connected both buildings. Although the only historical drawing of Dawes' observatory is a sketch included in his 30 April 1788 letter to Maskelyne, the building was carefully reconstructed in the now-defunct historical theme park, 'Old Sydney Town', near Gosford (New South Wales): see Figure 7 (bottom).

Dawes used the larger building to store his portable instruments, except for his thermometers, which were most likely stored in the well-ventilated observatory building (McAfee, 1981). The larger and more sensitive instruments, including his quadrant and the Shelton pendulum clock, were fixed in place. The historical record indicates that the quadrant was already put to practical use well before the observatory had been completed. It was clearly the most important instrument among the fledgling observatory's equipment, since it was used to accurately determine local time, longitude and latitude, and to calibrate the colony's clocks.

Dagelet referred to Dawes' quadrant in his letter of 3 March 1788, knowledge he most likely obtained following de Boutin's visit on 21 February: "I find that your q.c. [*quart de cercle*, quadrant] is perfectly well placed and leaves nothing to be desired from any point of view" (Morrison and Barko, 2009). The earliest observations with the quadrant at Sydney Cove may have been carried out by Lieutenant William Bradley (1757–1833) of the *Sirius*, rather than by Dawes (e.g., Johnson, 1998; Morrison and Barko, 2009): see Section 4 for details.

In his letter of 17 November 1788, Dawes informed Maskelyne about the arrangements he had made for the Shelton clock,

> On the 6$^{th}$ Sept.$^r$, I got the clock scrwed [*sic*] to the frame which is let into the niche in a very large solid stone and there wedged exceedingly firm and propped in front by a short, stout piece of plank to another very large stone so that I believe it is impossible to fix a clock up much better. (Dawes, 1788e)

Dawes expressed his satisfaction with the clock's firm foundation once again in a letter of 16 April 1790. He told Maskelyne that the clock had been positioned in a niche in a rock "which has never moved since the foundation of the world" (Dawes, 1790a).

**3.2 Routine operations**

Dawes devoted himself to "Observation", which encompassed taking standard astronomical, meteorological, magnetic and tidal measurements, drawing maps and recording the Eora language in collaboration with his Gadigal language partner, a young girl called Patyegarang (e.g., Nathan, 2009; Pybus, 2009; Gibson, 2010; Thomas, 2013). On 14 September 1788, he commenced routine astronomical and weather observations at his observatory, thereby taking on the role of the colony's unofficial meteorologist as well[15] (Russell, 1877: 2; McAfee, 1981; Orchiston, 1989; Derrick, 2019).

He took detailed astronomical readings and recorded temperature and weather observations—air pressure, wind strength and direction—between three and eight times a day until 6 December 1791 (Dawes, 1788–1791; Gergis et al., 2009). However, he only started recording daily rainfall measures, which he expressed in numbers of "grains", in September 1791, using a rain gauge constructed from a funnel and a "common quart bottle" (Ashcroft, 2016). Until 1 July 1791, Dawes also used a (faulty) barometer, whose readings were rendered inaccurate because of a crack in the cistern that allowed some of the mercury to escape:

> It was my intention to have it examined on its return to England, as, owing to the marker having depended upon glue to keep the cistern tight where it had been crack'd, it had but a small quantity of quicksilver [mercury], which I had discovered by observing a number of very small globules immediately under it on the floor of the observatory. In packing it up, however, I found a large crack in the cistern, thro' which perhaps all the quicksilver would have escaped in the course of the passage [to England] and have endangered breaking the tube; I therefore emptied it entirely and then packed it up. (McAfee, 1981: 21)

Dawes sent Maskelyne regular updates about his observational progress, although the variable weather presented a serious challenge: "there cannot be a more unfavourable country for observation" (Dawes, 1788e). In addition, Governor Phillip continued to order Dawes to take on ever more tasks in support of the fledgling colony, which eventually led to his frustrations getting the better of him. In a strongly worded letter written in July 1790, he told Maskelyne that "the very few men of business in this country have always each of them more to do than he can by any efforts perform properly", adding that "I do more real business than any besides [the chaplain] in the country" (Dawes, 1790b). Yet, as Governor, Phillip was ultimately in charge of the colony's day-to-day operations, and so Dawes had no choice but to act upon Phillip's practical orders—to the detriment of his scientific interests and achievements. This internal conflict may have exacerbated the men's apparently newly developed mutual antipathy, eventually leading to a complete breakdown of their personal relationship.

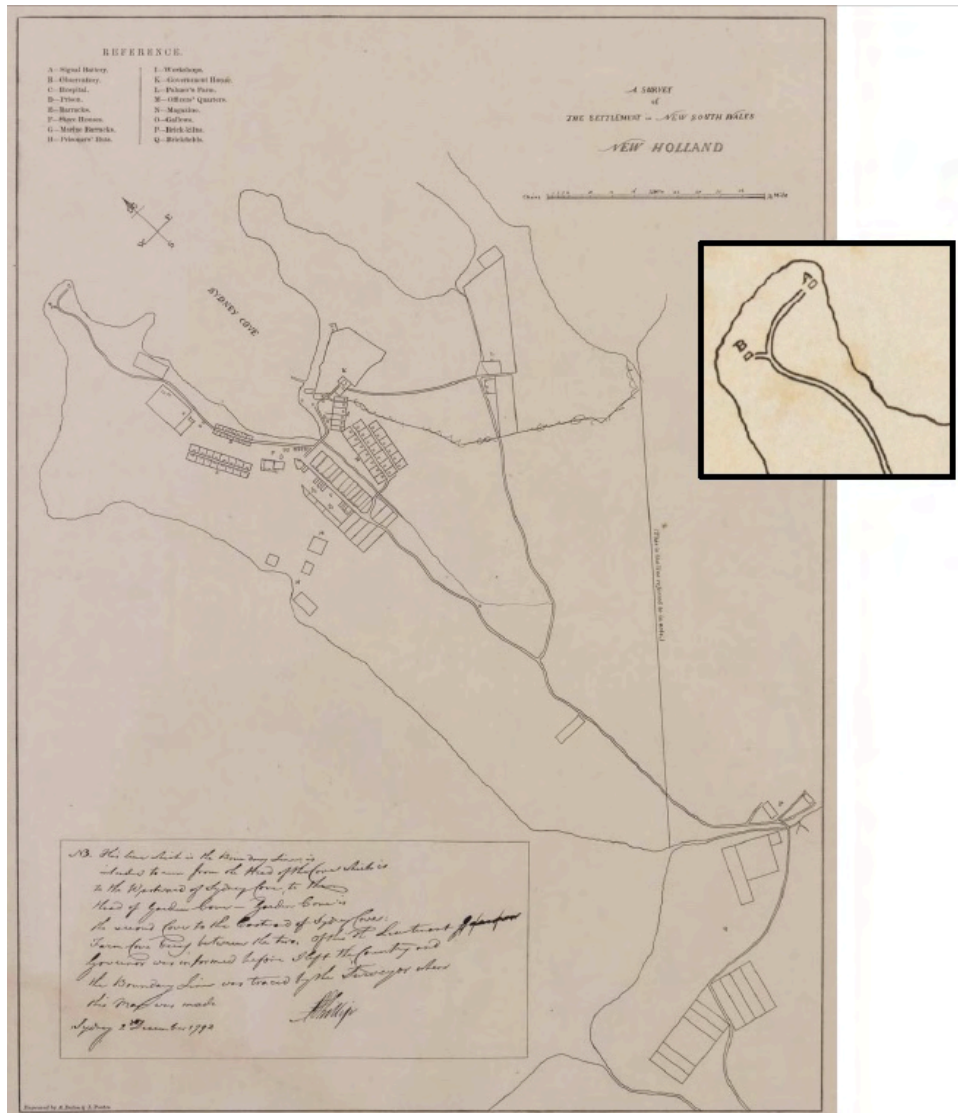

**Figure 8**: "A Survey of The Settlement in New South Wales, New Holland, by Lt. William Dawes, engraved by A. Dulon & L. Poates (Sydney, ca. 1886)." Dawes' observatory is building 'B'. For a high-resolution image, see https://nla.gov.au/nla.obj-229944353/ (National Library of Australia).

In addition to his duties as officer of the Marines and his roles as artillery officer and the colony's engineer, Dawes was tasked with designing the batteries for the colony's defence. Moreover, the alleged ineffectiveness of the settlement's Surveyor-General, August(us) Alt (1731–1815), significantly added to Dawes' workload since he was expected to come through as the settlement's *de facto* surveyor. His main and lasting achievements in this latter area—most likely undertaken with the astronomical quadrant as his principal tool—encompass his surveying and planning of the Government farm. In addition, he surveyed and planned the first streets and allotments in Sydney and Parramatta, the colony's second settlement, presently a major Sydney suburb (e.g., Haynes et al., 1996: 31–32). Figure 8, *A Survey of The Settlement in New South Wales, New Holland, by Lt. William Dawes, engraved* *by A. Dulon & L. Poates*, is the only remaining Crown plan (survey chart) of Sydney by Dawes himself (reproduced in 1886).

Most damningly, however, as well as unexpectedly indiscreetly, he complained to Maskelyne about Governor Phillip's apparent disinterest in the observatory's operations (e.g., Clarke, 2015). Nevertheless, he spent as much of his limited time as possible at the observatory, working late into the night, intent on finding Maskelyne's comet. As we learn from his letter of 17 November 1788,

> … the 7$^{th}$ of August I looked out all night for the comet ... and from that time to the present have not omitted looking for it at every available opportunity, but have not yet seen anything of it. (Dawes, 1788e)

Despite his failure to observe the comet, he discovered a number of new nebulae, observed a solar eclipse and took observations of the Moon's parallax and Jupiter's satellites (James, 2012). Indeed, from Mrs. Elizabeth Macarthur (1766–1850), the wife of John Macarthur (1767–1834), Lieutenant on the Second Fleet, we learn that Dawes was "so much engaged with the stars that to mortal eyes he [was] not always visible" (Macarthur-Onslow, 1973: 28)—this despite having agreed to act as her astronomy and botany instructor. Unfortunately, however, Dawes' observation records have been lost or misplaced. His letters to Maskelyne, although detailed and carefully documented, provide a mere summary of the full observation logs. Dawes' field journals were in the possession of the astronomer William Wales (1734?–1798), Commissioner of the Board of Longitude, at the time of the latter's death in 1798. They have not been seen since (e.g., Bosloper, 2010).

Dawes made his observatory into a welcoming place for everyone; it soon became the settlement's intellectual and cultural centre. Lieutenant Daniel Southwell (ca. 1764–1797), an officer from the *Sirius*, was clearly impressed: "He has a great share of genuine knowledge, studious yet ever cheerful and the goodness of his disposition renders him esteemed and respected by all who know him" (James, 2012).

Dawes realised, to his dismay, that the pendulum clock lost time during cold weather but gained time on hot days, while cloudy nights appeared to be the norm rather than the exception. In fact, in addition to his 'standard' astronomical duties, we learn from his correspondence with Maskelyne that he also intended to take gravity measurements using the pendulum clock (e.g., Morrison and Barko, 2009; Bosloper, 2010):

> I fix'd very firmly the Clock in a niche of a solid Rock, on Sat.$^y$ the 6th of Sept.$^r$ and by Altitudes taken at two or three days interval found it to be losing at the rate of about 36,"00 on sidereal time in one sidereal day but this Rate seems to be increasing as it did at Rio de Janeiro [where the First Fleet called into port on their voyage to New South Wales] for in these last 11 Days the Clock has lost after the Rate of 37,"25 on sidereal time in one sidereal day. (Dawes, 1788d)

Nevertheless, using the invariable, temperature-compensated grid-iron pendulum with a precisely determined pendulum length, he managed to compile a significant body of gravity measurements. These survive in his correspondence with Maskelyne (for details, see Bosloper, 2010). Gravity measurements involved a series of daily clock-rate readings, often aggregated over an entire month, which would be compared with astronomical time measurements, usually using the Sun's meridian passage.

**3.3 Change is afoot**

Meanwhile, Dawes continued his planning for the observatory's future. In the same letter of 10 November 1788, he advised Maskelyne that he hoped to find a suitable successor before his term in the colony came to an end. He even planned for the eventuality that he would fall seriously ill or die, in which case his close friend and the colony's chronicler, Watkin Tench, was on standby support:

> I have reason to believe that Captain Tench of the Marines, will in a moderate time become sufficiently acquainted with the practice of astronomy to be capable of supplying my place. (Dawes, 1788e)

Yet, by July 1789, only a year after Dawes' observatory had been established, it appears that the building had already become too small. Lieutenant Governor David Collins (1756–1810), Secretary to the Governor and Judge Advocate of the colony, wrote in his diary,

> The observatory building which was erected on our first landing being found small and inconvenient, as well as for the purpose of observing as for the residence of Lieutenant Dawes and the reception of the astronomical instruments, the stone-cutters began preparing stone to construct another, the materials for which were found in abundance upon the spot, the west point of the Cove. (Collins, 1798: 61)

A second, larger and possibly more durable observatory building was apparently under construction to replace the original timber and canvas structure. By April 1790, Dawes (1790a) proudly declared to Maskelyne that he had moved into a more comfortable house, while the Board of Longitude's instruments had been mounted securely, particularly the clock and the quadrant: "if could you see it... [you would agree that] it could not be better fixed". Johnson (1998) has pointed out that during construction of the new stone observatory, its wooden precursor most likely continued to see frequent use. Meanwhile, a small fortified battery and gun powder magazine were also under construction under Dawes' oversight, so that in mid-1789 there may have been a small cluster of buildings on Dawes Point.

As we have seen already, Dawes and Governor Phillip did not get along. Dawes felt that he was unreasonably held back in his astronomical aspirations, while for Phillip juggling the young settlement's safety, security and practical day-to-day operations were at the forefront of his mind. Their mutual dislike and disagreement came to a head by the end of 1790, when Phillip ordered Dawes and Tench to lead a punitive raid into the local Aboriginal community. Although Dawes eventually but reluctantly agreed to undertake the mission, he made it abundantly clear that he would not obey similar orders again. This was a clear breach of decorum, rightly interpreted as insubordination. A second infraction related to an illegal purchase of flour from a convict only made the situation worse.

Nevertheless, Dawes was keen to remain in the colony for an additional three-year term. However, even before Phillip formally notified him of his decision, Dawes expressed his doubts as to the potential success of his application: "I think it at least ten to one that I shall return with the Marine detachment" (Dawes, 1790b). And indeed, Phillip did not accept Dawes' reasons for his offending behaviour, advising him that the only way for him to remain in the colony was by unreservedly apologising for his misdemeanours while pledging to refrain from committing further infractions of a similar nature. Dawes refused. In December 1791, he left for England on the H.M.S. *Gorgon*, a 44-gun fifth-rate Royal Navy vessel in the Third Fleet.

Upon his departure from New South Wales, Dawes took with him the instruments, including the astronomical clocks, he had borrowed from the Board of Longitude. K1 returned to England in 1792 on the *Supply* (Hawkins, 1979).

Apparently, Dawes' back-up plan to have Tench step into the role of the colony's astronomer had not been activated. The building structure fell into disuse and disrepair; the observatory building collapsed. However, Collins (1798) reports that by the end of 1791 the rectangular wooden building had been appropriated as a guard room, a platform for a flagstaff, and that a cannon had been installed just behind it (Kerr, 1986). Sydney's scientific beginnings had come to a halt to give way to Britain's military might.

The last record of Dawes' observatory is found in the diary of John Crossley (1762–1817), the astronomer on board H.M.S. *Providence*, which called into Sydney on 28 August 1795 during her four-and-a-half-year round-the-world voyage known as the 'Vancouver Expedition':

I went on shore and examined the place where Mr. Dawes' observatory was built but found nothing standing but the uprights which supported the roof and the pillar on which he placed his quadrants. (James, 2012)

Nevertheless, detailed European maps and charts of Sydney Cove continued to include Dawes' observatory until at least 1798: see Figure 9 for a high-resolution example.

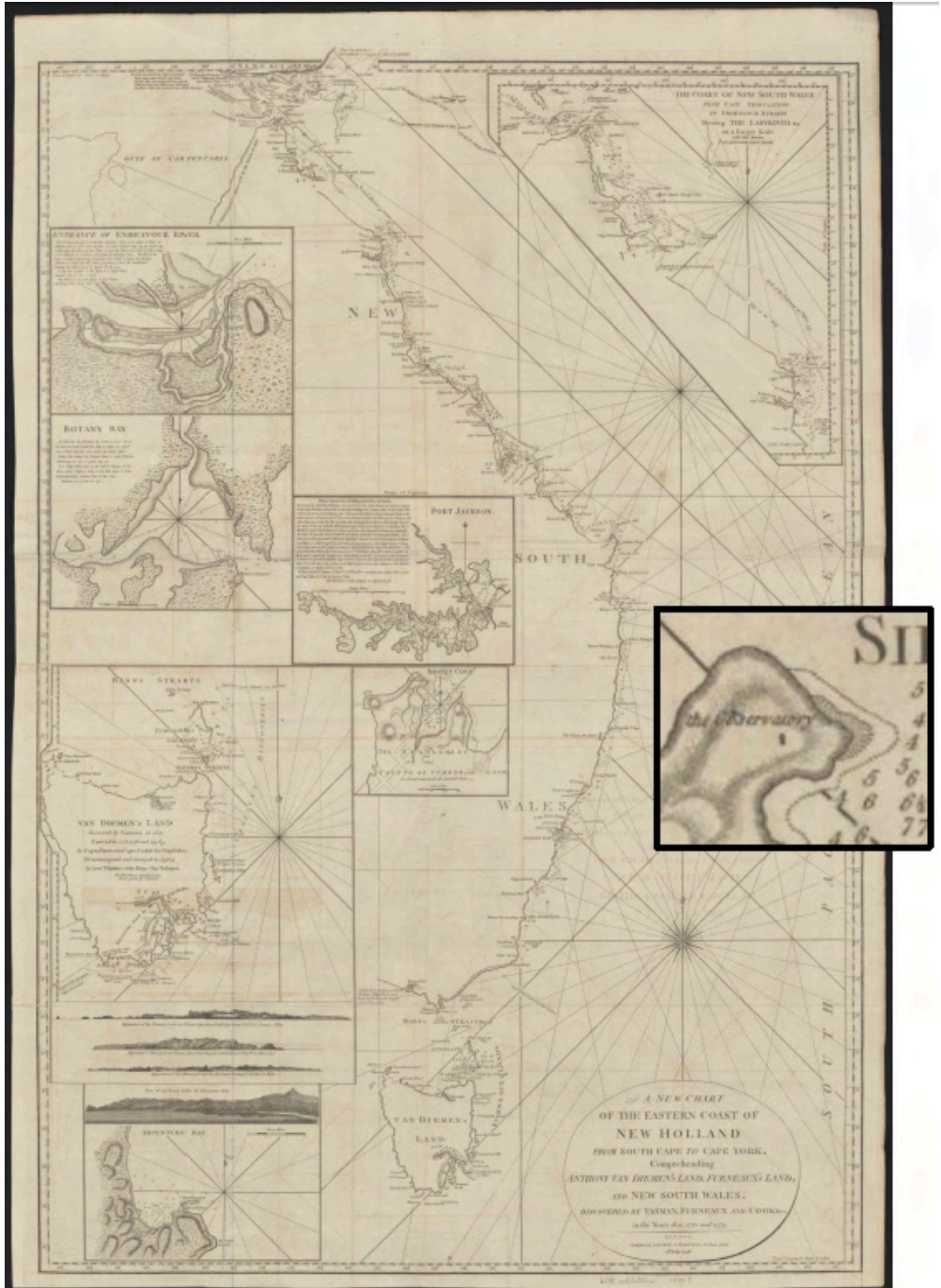

**Figure 9**: "A New Chart of the Eastern Coast of New Holland from South Cape to Cape York, Comprehending Anthony van Diemen's Land, Furneaux's Land, and New South Wales, discovered by Tasman, Furneaux and Cook &ca. in the Years 1642, 1770 and 1773. London, Published by Laurie & Whittle, 53 Fleet Street, 17th July 1798." For a high-resolution image, see https://nla.gov.au/nla.obj-368781110/ (page 105; National Library of Australia).

## 4 A MEMORIAL AT ODDS WITH THE HISTORICAL RECORD

The precise location of Dawes' observatory is still a matter of some debate. Here, we will shed further light on that debate. We aim to provide a comprehensive assessment of the most likely site.

### 4.1 The observatory's longitude and latitude

Let us start by offering an assessment of the various geographic longitude and latitude determinations for the observatory available in the historical record. McGuffie and Henderson-Sellers (2012) noted that all accounts, except for Dawes' (1788e) own determination, cite a latitude of 33° 52' 30" South, a location that is some 2.2 km south of the actual location. Dawes' own latitude determination is, however, quite precise; his geographic position determination is included as a mere postscript, an afterthought, in his letter to Maskelyne of 10 November 1788:

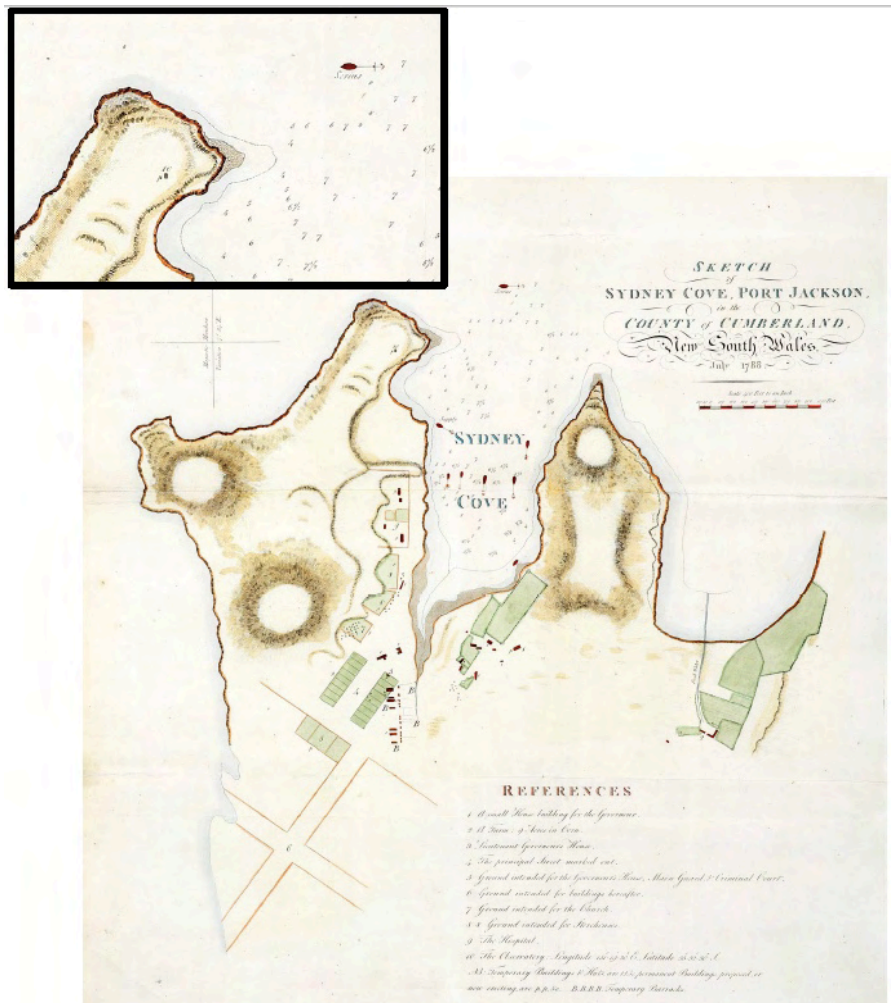

**Figure 10**: "Sketch of Sydney Cove, Port Jackson, in the County of Cumberland, New South Wales, July 1788. Captain John Hunter, William Dawes." (Source: William Dawes 1880; State Library of Victoria: ref. H24525.) For a high-resolution black-and-white copy, see
https://search.sl.nsw.gov.au/permalink/f/lg5tom/SLNSW_ALMA21145687720002626.

P.S. I have assumed the Longitude of the Observatory at 10h. 05 E. The Latitude I found roughly by the astronomical quadrant mounted on the stump of a tree saw'd off, in the open air to be 33.° 52' 20"; but by the zenith distances accompanying this I take it to be about 33.° 51' 10" … The longitude I have assumed is what Capt.$^n$ Hunter & M.$^r$ Bradley made it by a mean of upwards of 300 Distances. (Dawes, 1788e)

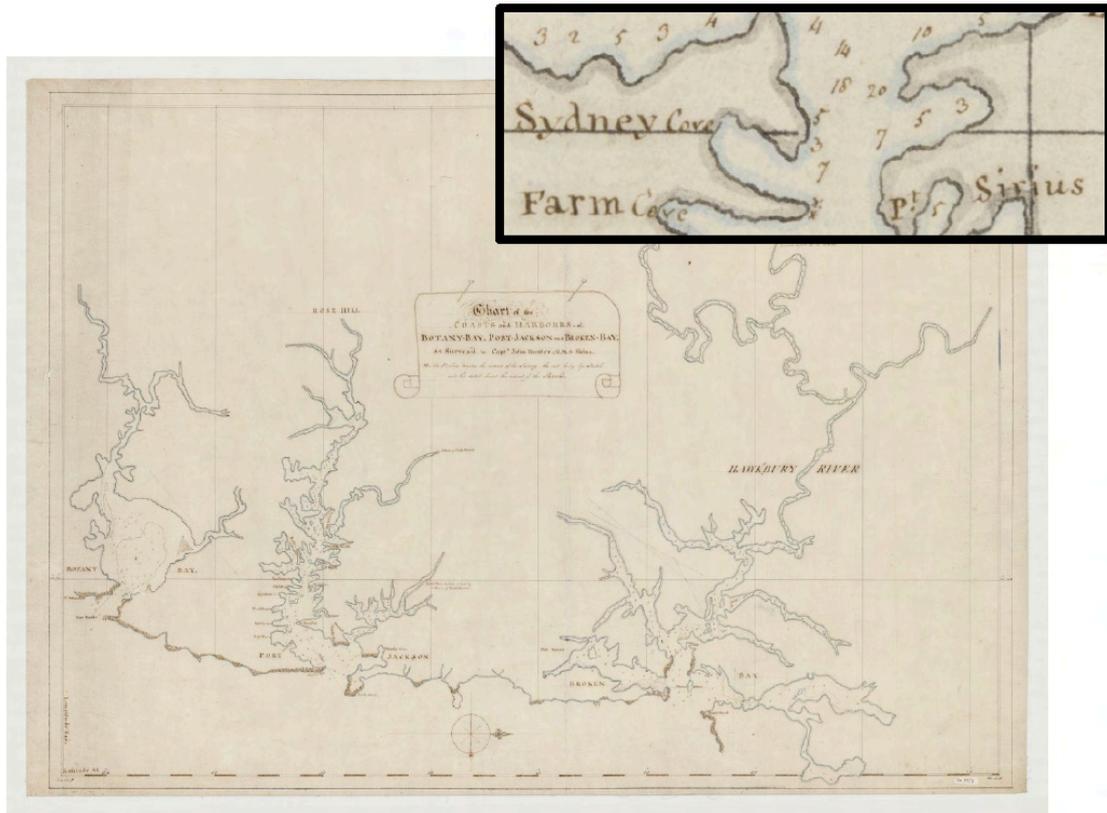

**Figure 11**: "Chart of the Coasts and Harbours of Botany-Bay, Port Jackson and Broken Bay on the coast of New South Wales as Survey'd by Capt.[n] John Hunter of H.M.S. *Sirius* (1789?)." For a high-resolution image, see https://search.sl.nsw.gov.au/permalink/f/lg5tom/SLNSW_ALMA21147000280002626 (Mitchell Library, State Library of New South Wales).

Morrison and Barko (2009) suggest that Dawes' measurements were obtained as early as February or March 1788; they certainly predate the establishment of Dawes' observatory given that the quadrant was positioned on what is likely to have been a substantial "stump of a tree saw'd off" that could support the instrument's solid cast-iron base—and not on the bedrock where Dawes would eventually fix it firmly in place. The historical record includes a second measurement by Dawes, in the form of a carefully drawn sketch map of Sydney Cove published in July 1788, with Hunter, as part of *A Voyage of Governor Phillip to Botany Bay* (London, 1789; see Figure 10). The latitude included on this map, 33° 52' 30" South, is slightly less accurate, although the longitude cited is somewhat more accurate (see below). It is, therefore, likely that Dawes and Hunter determined their geographic positions independently.

As indicated, Dawes had adopted the longitude determinations of Bradley and Hunter. By the end of April 1788, Bradley had independently determined the observatory's latitude at 33° 52' 30" South based on "3 meridian altitudes of the Sun with the astronomical Quadrant"; his longitude measurement, based on 176 lunar distances, resulted in 151° 20' East (Dawes, 1788e; Bradley, 1969). Bradley's *Chart of the Coasts and Harbours of Botany-Bay, Port Jackson and Broken Bay on the coast of New South Wales as Survey'd by Capt.[n] John Hunter of H.M.S. Sirius* (1789?)[16] includes a single meridian, the first Australian meridian published, determined at Dawes' observatory (see Figure 11, inset). Hunter took 130 lunar distances and obtained a longitude within a mile of Bradley's result (Dawes, 1788e).

It is unknown what type of observations Dawes obtained himself at this time (see also Dawes, 1788a). Dawes' own map of March 1791, *A map of all those parts of the territory of New South Wales which have been seen by any person belonging to the settlement established at Port Jackson... / faithfully constructed ... and respectfully inscribed to Capt*

*Twiss... By ... William Dawes*, is the only other contemporary map that includes a meridian: see Figure 12. Dawes included an explicit note regarding his meridian, whose extension also appears to go through the location of Dawes' observatory: "N.B. The longitude of the observatory is determined from the observations of Capt. Hunter and Lieu.[t] Bradley".

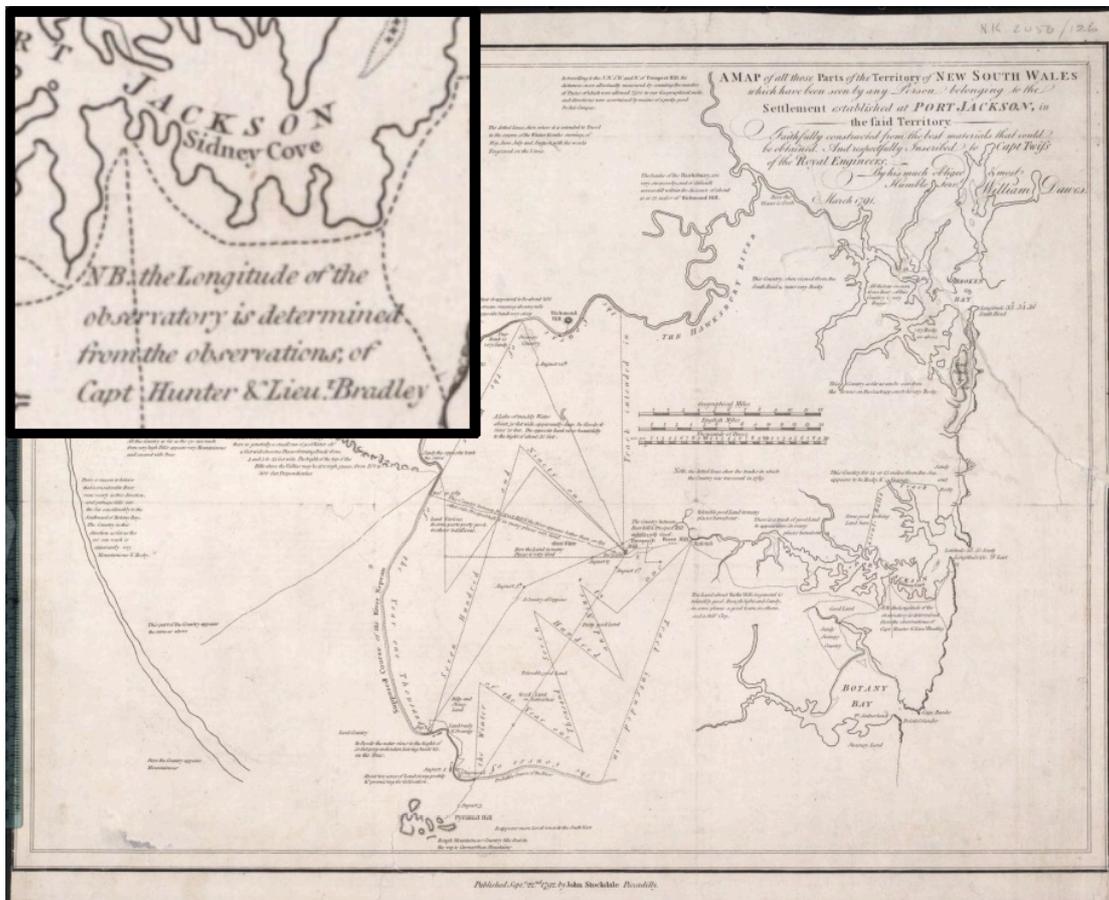

**Figure 12**: "A map of all those parts of the territory of New South Wales which have been seen by any person belonging to the settlement established at Port Jackson... / faithfully constructed ... and respectfully inscribed to Capt. Twiss... by ... William Dawes, March 1791." For a high-resolution image, see https://nla.gov.au/nla.obj-230635598/ (National Library of Australia).

Meanwhile, Phillip's journal includes a position determination for the observatory that is almost certainly a transcription error: "The longitude of this observatory is ascertained to be 159°19' 30" east from Greenwich" (Phillip, 1790). This is most likely meant to be 151° 19' 30" East, since Collins' (1798: 15) published longitude for the observatory, included as a journal entry for February 1788, is 151° 19' 30" East, as is the longitude cited on the Dawes and Hunter map of July 1788. Dawes' own measurement, 10 hours and 5 minutes East of Greenwich, translates to a longitude of 151° 21' East. Table 1 includes an overview of all contemporary geographic position determinations of Dawes' Observatory, approximately in order of publication.

**Table 1**: Contemporary geographic position determinations of Dawes' Observatory.

| Source | Latitude (South) | Longitude (East) | Notes |
|---|---|---|---|
| Bradley (1969) | 33° 52' 30" | 151° 20' | (a) |
| Dawes (1788e) | 33° 52' 20" | 151° 21' | (b) |
|  | 33° 51' 10" |  | (c) |
| **Dawes and Hunter (1788)** | **33° 52' 30"** | **151° 19' 30"** |  |
| Phillip (1790) |  | 159° 19' 30" | (d) |
| Collins (1798) |  | 151° 19' 30" |  |
| Google (2020) | 33° 51' 18" | 151° 12' 35" |  |

***Notes***: (a) Most accounts (McGuffie and Henderson-Sellers, 2012); (b) Based on measurements with the quadrant; (c) Based on zenith distance measurements; (d) '159' is likely an erroneous transcription (see text).

The most popular contemporary geographic position determination, that of Dawes and Hunter (1788; rendered in bold italic font in Table 1), places the observatory some 11 km east of the actual position on Dawes Point/The Rocks and around 3.8 km off the coast of New South Wales (McGuffie and Henderson-Sellers, 2012). At the latitude of Sydney, 33° 52' 08" South, an East–West discrepancy of 11 km corresponds to 7.1 minutes of arc in longitude (the actual longitude difference between the Dawes and Hunter determination and the modern value is 6'55"), or a timing error of 29 seconds. Contemporary time measurements, including time differentials with respect to a reference meridian, were much more accurate than this, however (e.g., de Grijs, 2017). Morrison and Barko (2009) suggest that this positional mismatch may have been a systematic offset caused either by incorrect lunar-distance and astronomical ephemeris tables or by problems with the K1 chronometer since it had been allowed to run down during the Indian Ocean crossing. However, the geographic location of the French tent observatory's site on the shores of Botany Bay determined by Dagelet is similarly displaced from its modern position (Morrison and Barko, 2009). This suggests that the positional mismatch was less affected by the (in)accuracy of K1 and more so by the problems related to the accuracy of the contemporary almanac tables. Both Dagelet and Dawes would most likely have been aware of these issues.

**4.2 Present-day context**

At the present time, the headland to the west of Sydney Cove includes the Sydney suburbs of The Rocks on the eastern side and Dawes Point on the western side of the promontory's ridge. The area of historical interest is no longer contained by the current boundary of the Dawes Point suburb, which was formally established in 1993 (Jackson, 2018b). Since Dawes' return to London in 1791, the area known as Dawes Point gradually moved (north-)westwards from the original 'Point Maskelyne' to the present-day location of the Pier One hotel and Ives Steps at Walsh Bay. Except for Dawes' and Hunter's July 1788 map showing the entire headland, there are no historical records of the astronomer having been active on the western side of the area.

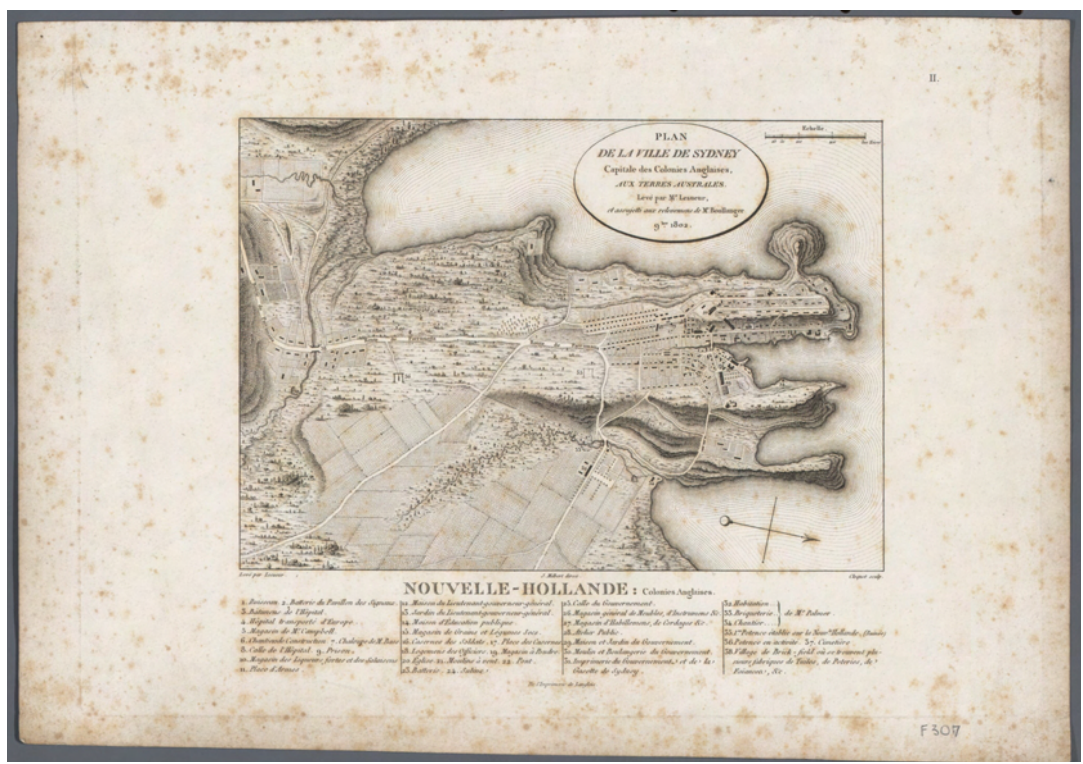

**Figure 13**: French map of Sydney, 1802. For a high-resolution image, see https://nla.gov.au/nla.obj-229944462/ (National Library of Australia).

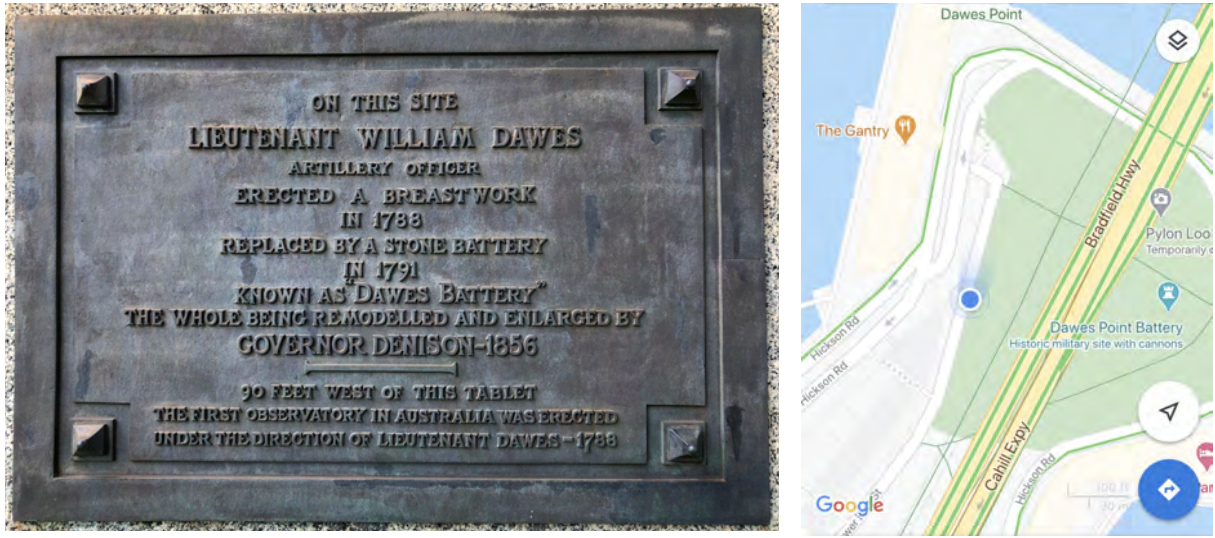

**Figure 14**: (*left*) Dawes' memorial plaque attached to the Sydney Harbour Bridge (Photo: Richard de Grijs). (*right*) Blue dot: location of Dawes' observatory implied by the text on the plaque.

The shifting boundaries of the Dawes Point area coincided with the developing fortifications at the site (Jackson, 2018a), *viz.* Dawes' battery (closed in 1916) and Fort Phillip (present-day Sydney Observatory, established in 1858). The southern pylon of the Sydney Harbour Bridge is located at the northern extremity of the promontory. The area coincident with the location of Dawes' observatory, today's Dawes Point (Tar-ra) Park, part of the Hickson Road/Dawes Point Reserve, is located northwest of a small inlet known as Campbells Cove, near the present-day Park Hyatt Hotel.

Nevertheless, a number of contradictory identifications of the observatory's location appear in the literature. Johnson (1994), archaeologist with the Sydney Cove Authority (New South Wales Government), declared that Dawes' observatory would have been located on the promontory's highest point, a conclusion echoed by Jackson (2018a) and Johnson (2003). The latter publication implies that a single building located on the promontory's ridge and shown in a drawing from the early 1790s (Johnson, 2003; his Figure 2) may have been Dawes' observatory. However, this view is at odds with Dawes' own description of the observatory and its site, including the dual-level configuration of the structure. Wood's (1924) suggestion that the *Sirius*' guns may have been located *in front of* the observatory (see note 9), with respect to the waterline, may also have contributed to the confusion. Contemporary maps of Dawes Battery (for a typical example, see Figure 13) show the gun placements close to the waterline, in front of a steep incline.

Johnson (1994) refers specifically to a bronze plaque, located on Pier No. 6, the first pier 150 feet (46 m) southwest of the southern pylon of the Sydney Harbour Bridge (*Daily Telegraph*, 1926; *Sydney Morning Herald*, 1932), as his position reference. As we will see below, this South-facing tablet implies that the original observatory was located some 90 feet (27 m) to the west of the pillar of interest: see Figure 14 (left). Between February and September 1995, the Sydney Cove Authority undertook archaeological excavations near the area suggested by Johnson's (1994) report, although without locating any remains of the observatory. Johnson (2003) suggested that this failure could be related to the site's subsequent use as stone quarry around 1819. However, as we will show below, the memorial plaque's indicative position of Dawes' observatory (see Figure 14, right) is most likely incorrect. In addition, Collins (1798: 61) implies that Dawes and his men were quarrying stone for the second incarnation of the observatory, suggesting that the observatory's site was not coincident with but somewhat removed from the main quarry location.

To uncover the full extent of the historical record, let us first briefly summarise what we have learnt from contemporary maps and Crown plans. The majority of contemporary maps showing Dawes' observatory were drawn by either Hunter or Dawes himself, including Hunter's first map of the settlement dated 1 March 1788 (Figure 5), a joint map by Hunter and Dawes from July 1788 (Figure 10), Hunter and Bradley's chart of the three harbours in New South Wales (Figure 11), Dawes' chart of the greater Port Jackson area of March 1791 (Figure 12) and Dawes' contemporary Crown plan, which was however only published by 1886 (Figure 8). This latter map is a tracing based on Crown Plan CP1-172. It shows the observatory ('B') and the signal battery ('A'). The observatory is sited on the western side of Dawes Point. This is in conflict with all other early and contemporary maps, unless A and B have been transposed, which we suspect may have happened. We inspected the original Crown Plan at the State Archives of New South Wales, where we noticed that this particular map is severely damaged at the location where its key would have been located (the key is no longer present). The sketch map of Figure 10 is the most carefully drawn contemporary map, although it does not scale to modern shorelines. The observatory is marked '*p*': 'proposed or now building'.

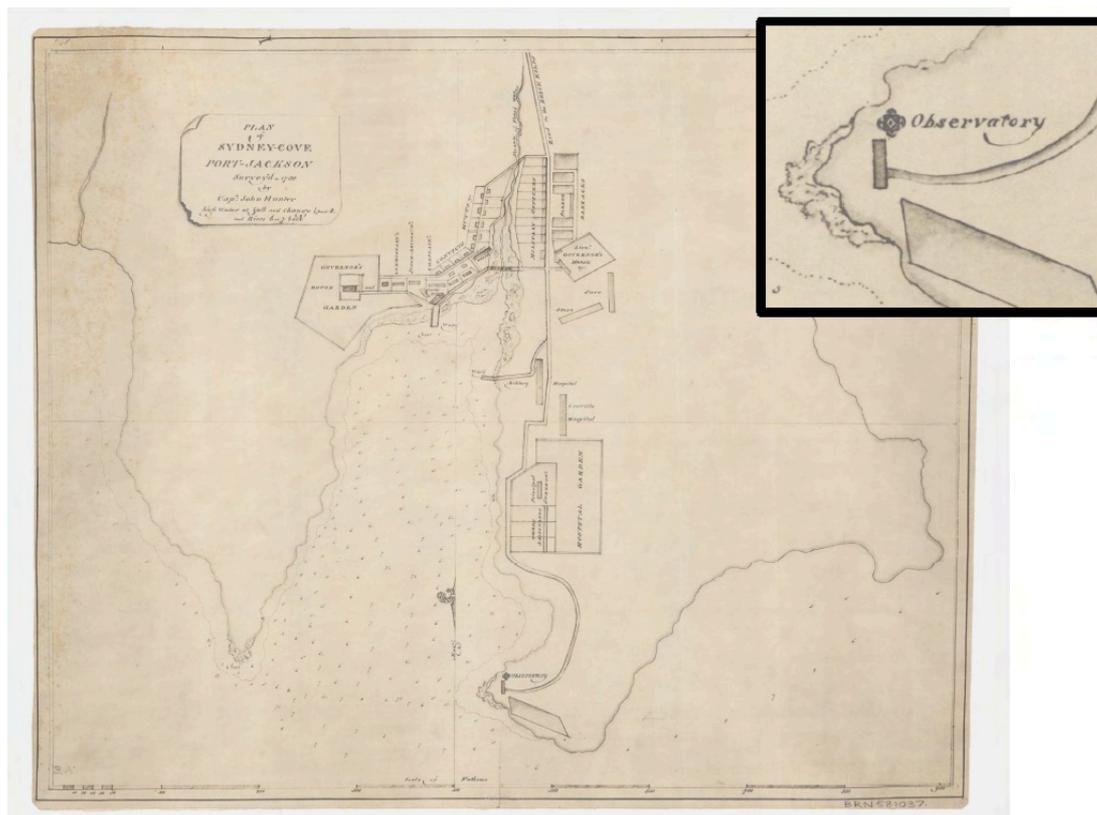

**Figure 15**: Plan of Sydney-Cove, Port-Jackson survey'd in 1788 by Capn. John Hunter. For a high-resolution image, see https://search.sl.nsw.gov.au/permalink/f/lg5tom/SLNSW_ALMA21138514930002626 (Mitchell Library, State Library of New South Wales).

We have uncovered two additional contemporary maps and charts that also show the location of Dawes' observatory, a survey map by Hunter from 1788 (see Figure 15: note the dual-building structure shown for the observatory; this is likely an intended plan rather than a record, also because the Governor's residence appears out of place) and a convict's drawing of the Sydney Cove area dated 16 April 1788 (Figure 16). This latter map is usually attributed to Francis Fowkes (fl. ca. 1788–ca. 1800), who is possibly identified with the 'Port Jackson Painter' whose sketches of the early colony continue to provide valuable insights into the pre-colonial era. The latter drawing, although more conceptual than geographically correct, is important, since it represents independent evidence.

All contemporary maps and charts show the site of Dawes' observatory on the eastern side of the headland, where indicated situated against a rocky bluff. Figure 17 provides an overview of the locations implied by the historical maps and charts discussed in this article. Panels (a) through (c) show representative examples of contemporary maps overlaid on a modern outline of Sydney Cove, including (a) Hunter's first survey map of Sydney Cove of 1 March 1788 (Figure 5), (b) his planned survey of 1788 (Figure 15) and (c)

Hunter and Dawes' careful sketch from July 1788 (Figure 10). Figure 17d shows the positions of Dawes' observatory (and the meridian based on the observatory's fixed quadrant) according to the maps and charts discussed in this article. Note that while we have attempted to reconcile Dawes' survey map, Figure 8, with the modern map, the match is tentative at best since the contemporary survey's scaling of identifying features varies across the original map.

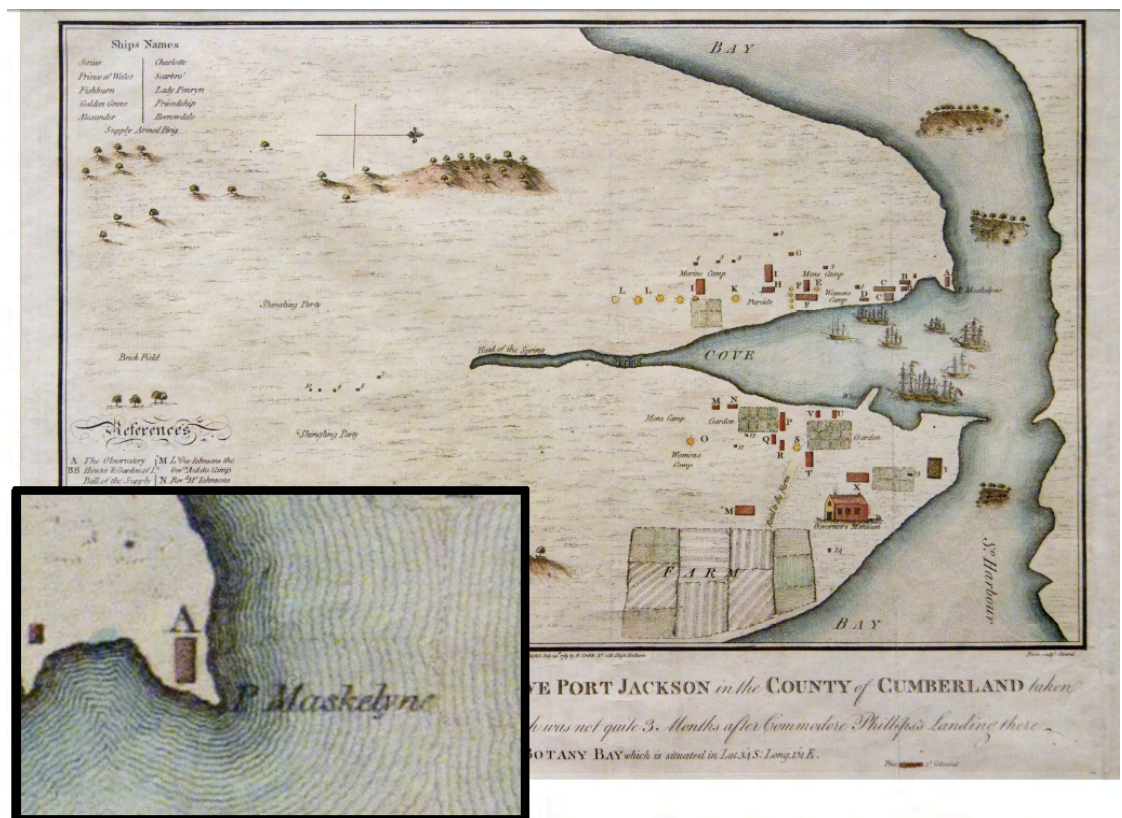

**Figure 16**: Sketch & description of the settlement at Sydney Cove, Port Jackson, in the County of Cumberland taken by a transported convict on the 16th of April 1788, which was not quite 3 months after Commodore Phillips's landing there. For a high-resolution image, see https://search.sl.nsw.gov.au/permalink/f/s8mhc3/SLNSW_ALMA211126408850002626 (Mitchell Library, State Library of New South Wales).

To further support our conclusion of the observatory's location on the eastern side of the promontory, in Figure 18a we provide a wide-angle view of the site's present-day appearance. The memorial plaque is visible on the pier on the left of the photograph. We have indicated the area covered by Dawes' Battery by means of a light green outline, a location supported by both archaeological records and today's restored battery. Beyond the battery, the yellow arc traces the edge of the bedrock; the yellow arrow indicates a solitary cannon at that position. As we have seen, Dawes' letters to Maskelyne imply that he built his observatory against the edge of the bedrock. Projection effects (foreshortening) render a proper assessment of the distance (~20–30 m) between Dawes' Battery and the edge of the bedrock difficult, however. Therefore, Figure 18b shows an enlarged image of the same area but from a different vantage point, clearly showing this distance.

We are fairly confident in our determination of the most likely observatory site, because (i) it would not be prudent to build an observatory by the shoreline; (ii) Dawes' (1788a) letter to Maskelyne implies that he constructed the rectangular building against a large block of rock and the octagonal building on top of it; (iii) we have established that the facility was built on the eastern side of the headland; and (iv) it was likely built close to the stone quarry later identified by Johnson (2003). However, we caution the reader that we do not have enough physical or documentary evidence to state categorically exactly where the building was located. Any physical evidence has been erased by the later fortification work (which likely remodelled the hillside) and the construction of the Harbour Bridge (see below).

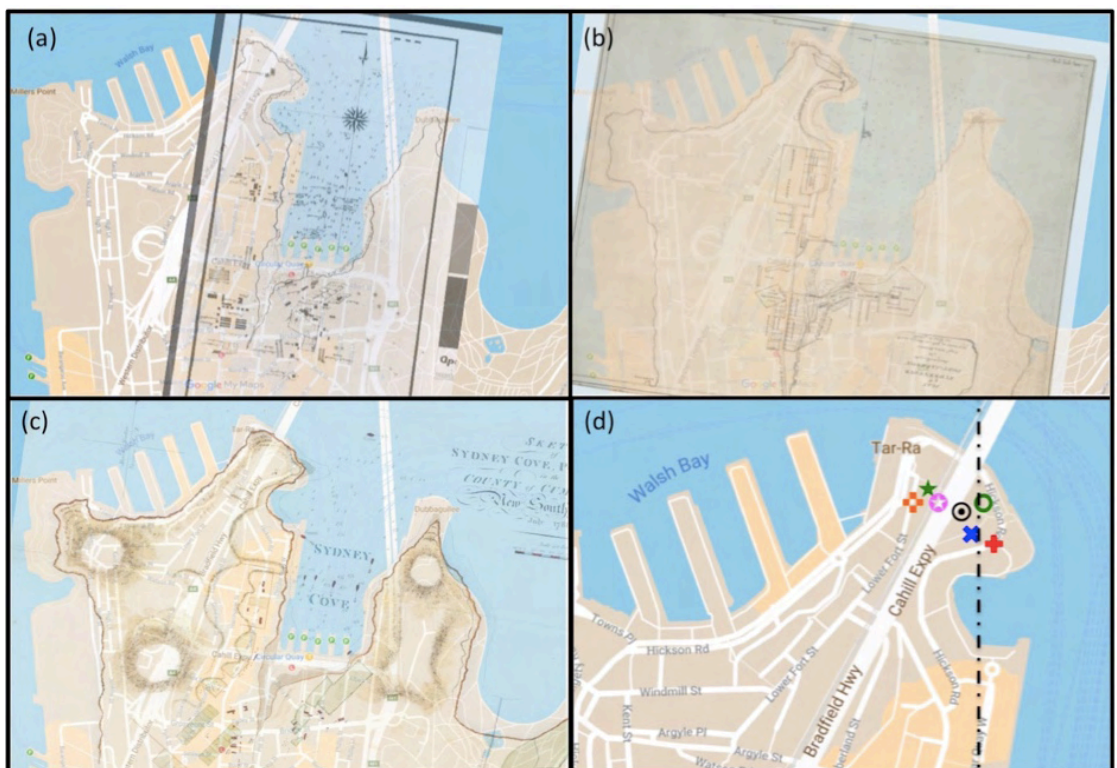

**Figure 17**: (*a*), (*b*) and (*c*) Examples of matching the contemporary record (Figures 5, 15 and 10, respectively) with a modern map (© Google, 2020; permissible use; each panel is 1,900 × 1,400 m$^2$). (*d*) Locations of Dawes' observatory implied by the records discussed in this article. *Black* ⊙: Hunter's map of 1 March 1788 (Figure 5); *red* ✚: Hunter's 1788 planned survey map (Figure 15); *blue* ✖: Hunter and Dawes' 1788 map of Sydney Cove (Figure 10); *dark green*: Locations of the observatory ('B'; ★) and Dawes' Battery ('A'; ⭘) according to Dawes' Crown Plan (Figure 8); *orange* ✣: Location indicated on the Sydney Harbour Bridge's memorial plaque; *pink* ⊙: Approximate location of the archaeological excavations described by Johnson (2003). The black dash-dotted line corresponds to the best match to the meridian line shown in Figure 11. Panel *(d)* covers an area of approximately 850 × 650 m$^2$. The convict's sketch shown in Figure 16 is not directly transposable.

Yet, the evidence of a site on the eastern side of the promontory appears overwhelming, and so we were rather perplexed by the indication of a western site on the memorial plaque attached to the Sydney Harbour Bridge. Therefore, we decided to explore the origin of the citation on that tablet.

The memorial plaque was formally unveiled on 16 June 1932 (*Sydney Morning Herald*, 1932), but the idea to establish a plaque as a memorial of the local history goes back to a letter from John Job Crew Bradfield (1867–1943), Chief Engineer of the Sydney Harbour Bridge, to the Secretary of the RAHS, Karl Reginald Cramp (1878–1956), of 23 March 1926: see Figure 18. Bradfield advised the RAHS that "the Minister of Public Works[17] wishes to commemorate this historical battery [Dawes' Battery] and desires a brass plaque suitably inscribed to be erected" (Bradfield, 1926). He proceeded to ask the Society to provide a suitable inscription. In the RAHS Annual Report of 1926 (RAHS, 1927), we learn that the President and Secretary of the Society, Captain James Henry Watson (1841–1934) and Cramp, reached agreement on the proposed wording (see also *Daily Telegraph*, 1926; *Sydney Morning Herald*, 1927; RAHS, 1928 and 1931):

> At Dr. Bradfield's request, a suitable inscription for a brass tablet to be affixed to one of the granite-faced piers of the North Shore Bridge [the Sydney Harbour Bridge], was suggested by your President and Secretary. The proposed inscription is worded as follows:–
>
> On this site
> Lieutenant William Dawes (Artillery Officer)
> erected a breastwork in 1788,
> replaced by a stone battery in 1791,

known as Dawes Battery,
the whole being remodelled and enlarged by
Governor Denison, 1856.
**— feet[18] west of this tablet**,
the first Observatory in Australia
was erected
under the direction of Lieutenant Dawes,
1788

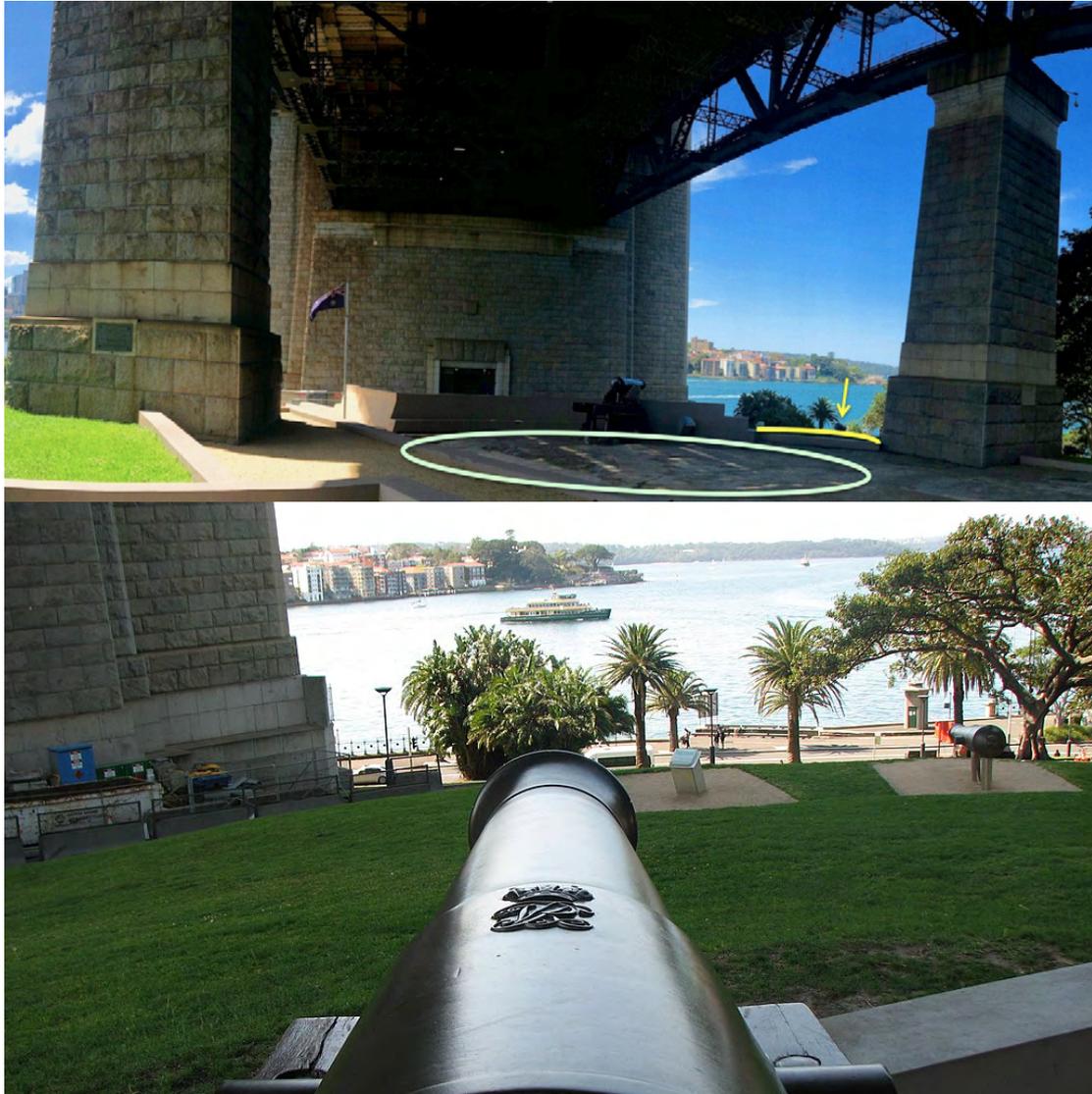

**Figure 18**: (*a*) Today's appearance of Dawes' Battery under the southern approach to the Sydney Harbour Bridge. The memorial plaque is visible on the pier on the left of the photograph. The light green outline reflects the location of the original battery, while the yellow arc traces the edge of the bedrock. The yellow arrow indicates a solitary cannon at the edge of the bedrock. (Photo: Richard de Grijs) (b) Overview of the area of interest, highlighting the non-negligible distance between the location of Dawes' Battery, from which this photo was taken, and the edge of the bedrock (at the location of the cannon at top right), where the astronomer established his observatory (Credit: Collywolly, Wikimedia Commons; Creative Commons Attribution-Share Alike 4.0 International license).

It is thus clear that the original wording proposed by Watson and Cramp (the emphasis is ours) was already problematic. Unfortunately, we have not been able to find further details as to the origin of the mistake. The Mitchell Library, part of the State Library of New South Wales, contains some of Watson's personal correspondence covering this period. However, none of those letters shed any light on the origin of the citation on the memorial plaque. Similarly, Figure 19 represents the only correspondence regarding the Dawes' memorial tablet contained in the RAHS Library (D. Newton, personal communication).

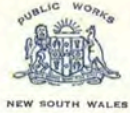

**Figure 19**: Letter from Chief Engineer Bradfield to the RAHS Secretary, 23 March 1926 (Courtesy of the RAHS Library; reproduced with permission).

Note that Bradfield's letter to the RAHS states that "during the construction of the Approach Spans [now known as Bradfield Highway] of the Sydney Harbour Bridge it was necessary to demolish the old battery at Dawes Point …" Indeed, Figure 20 shows one of two enormous pits—40 feet (12.2 m) wide, 90 feet (27.4 m) long and up to 30 feet (9.1 m) deep—known as 'skewbacks' at the southern end of the Sydney Harbour Bridge span (NSW Education Standards Authority, 2014). Their excavation, in preparation of the Harbour Bridge's pylon construction, may have contributed to the removal of any remains of Dawes' Battery and his observatory. Although the skewbacks did not extend as far from the southern end of the headland as the likely location of Dawes' Observatory, other contemporary photographs in the Mitchell Library's collection clearly show that the entire area under the southern approach to the Harbour Bridge was deeply excavated and thoroughly disturbed during the Bridge's construction. Moreover, one should keep in mind that Dawes' observatory, a feeble wooden structure constructed during the first year of British settlement, was unlikely built on deep and solid foundations. Without access to proper tools and professional construction expertise, even the stone building that was under construction just a year later was likely still a simple structure without deep foundations, given the thin layer of top soil on bedrock that was so prevalent around Sydney Cove (e.g., Hughes, 1987: 107). It is therefore not surprising that the 1995 archaeological excavation (Johnson, 2003) was not successful in its quest to locate any remains of Dawes' observatory.

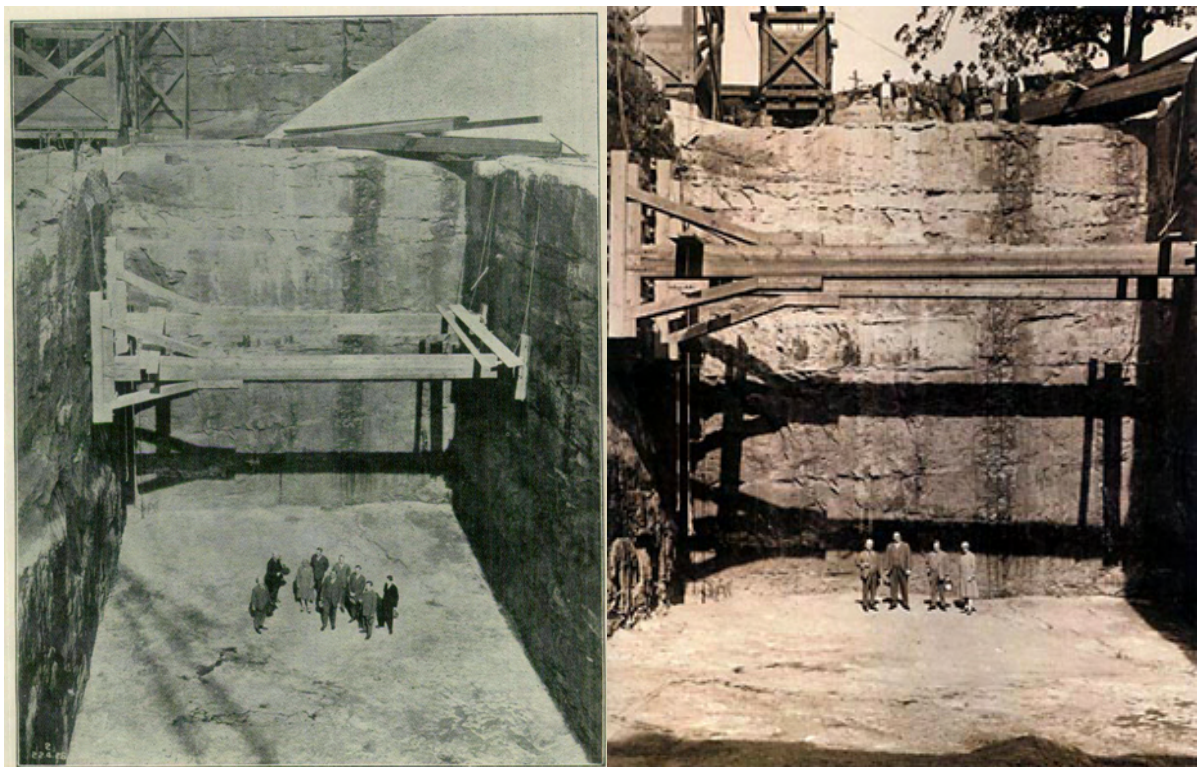

**Figure 20**: (*left*) Sydney Harbour Bridge, excavation of the South pylon's southeast 'skewback'. (Department of Public Works, 1927: 57) (*right*) Harbour Bridge Chief Engineer John Bradfield, construction supervisor Lawrence Ennis, structural engineer Ralph Freeman and Kathleen Butler, technical expert representing the State Government, in the excavation for the southwest skewback, 22 April 1926 (*Sydney Harbour Bridge Photographic Albums*, 1923–1933, Vol. 2; State Records of New South Wales).

## 5 SYDNEY'S SCIENTIFIC BEGINNINGS: CONCLUSION

One aspect we have not yet considered is the evidence provided by the morphology of the landscape. Let us therefore explore this latter aspect as additional, although perhaps less conclusive, evidence. While landscape artists usually tried to reproduce an area's main features accurately, some measure of creative license often crept in. In this context, we are most interested in exploring the shape of the original headland, and particularly the location of any obvious vertical rock faces. We have to cast our net more widely than simply focusing on the period from 1788 to 1791, however.

The Sydney harbourside topography is shaped by the nature of the so-called Sydney Basin Hawkesbury sandstone it is composed of. The rock strata is horizontal in the harbour region, and weathering causes horizontal and vertical fractures. This results in the topography of low, stepped headlands, with a short shoreline cliff backed by an approximately horizontal—sometimes gently sloping—terrace followed by one or more additional vertical terraces and steps. Higher headlands rise in a steeper set of terraces and steps up to a plateau. In reality, this terrace and step topography tends to be more chaotic and broken. From a distance, the features can be disguised and 'rounded' by vegetation. At the shoreline, a broad tidal terrace is common. Today, most shorelines have been modified by quarrying or infill behind rock retaining walls, although the natural form can still be seen at Ball's Head and in large parts of Cremorne Point and Bradley's Head.

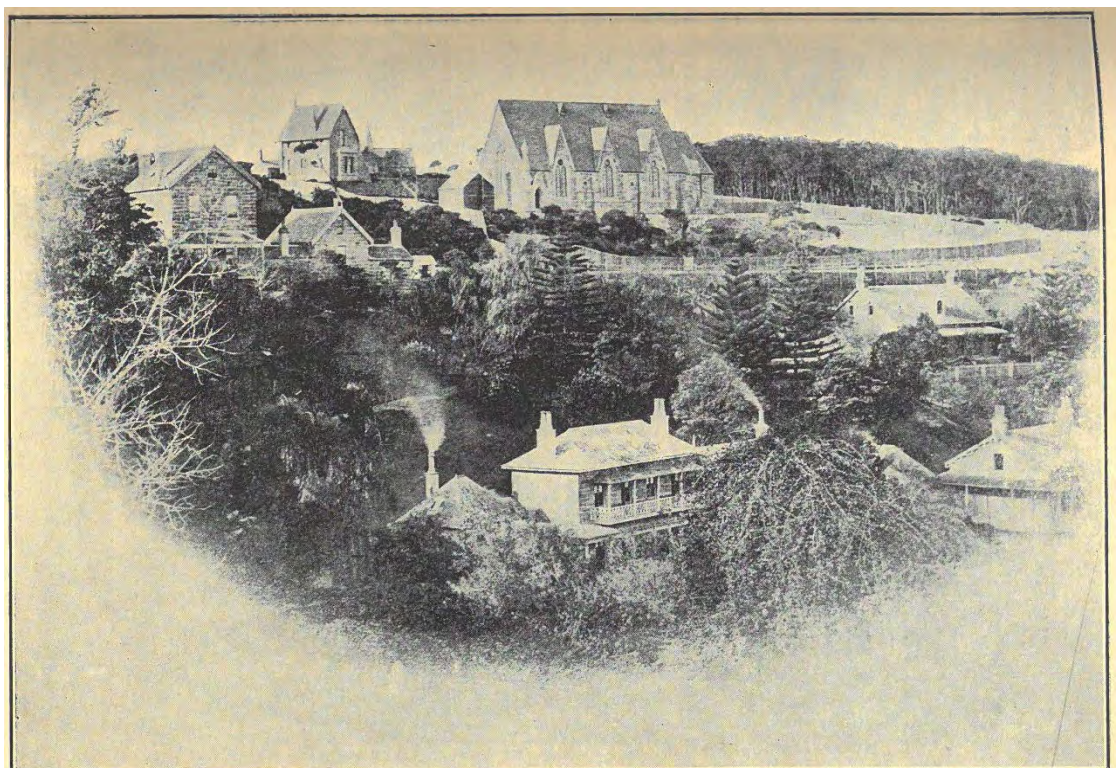

**Figure 21**: View of Lavender Bay in North Sydney (reversed) prior to the construction of the Sydney Harbour Bridge. The 'fuzzy' tree on the hill at the image's top left marks the site where the Bridge's first sod was turned on 28 July 1923 (Cash, 1930: 1).

Figure 21 shows the original morphology of the harbour shore at Lavender Bay (the image is reversed with respect to the original), near the start of the Sydney Harbour Bridge but prior to the start of construction. As the author intimates, the site of the 'fuzzy' tree on the hill at the left of the image is where the first sod was turned on 28 July 1923, marking the start of construction a few days later (Cash, 1930: 7–8). The age of the picture is uncertain, other than that the author wistfully refers to the "old-time picturesqueness" of 60 years ago, that is, around 1870. He proceeds (our emphasis),

> One afternoon, some years ago, I met my friend, Mr. James Blue, of honourable memory. He said, "Come inside and I will show you something that will interest you". … There came into his possession some time ago, a photograph, yellow with age, but alive in historic interest. (Cash, 1930: 9)

Our first, roughly contemporary sketch of interest is titled *A View of Dawes Battery at the Entrance of Sydney Cove*: see Figure 22. The sketch was published in 1820 by Walter Preston (ca. 1787–after 1821), a convict engraver (Willetts, 2010–2020). It was based on an original drawing by Captain James Wallis (1785?–1858), commander of the convict settlement in Newcastle (New South Wales). Wallis was based in Sydney between February 1814 and June 1816. The drawing clearly shows a low-lying extension to the headland, which rises steeply to a higher-level, isolated building. This same morphology of the bedrock is confirmed in a detailed sketch by the artist and former convict John Eyre (1771–1812; Figure 23), which is thought to have been created around 1809. Eyre's sketch is particularly valuable, because he carefully sketched the shapes of the buildings individually rather than using generic shapes. In addition, the contemporary oil painting from ca. 1800 (artist unknown) shown in Figure 24 also implies a similar form of the headland.

All of these early representations viewed Dawes Point from either the East or the North East. Before concluding this section, we highlight two additional paintings, shown side-by-side in Figure 25. Both view Dawes Point from due North, and both suggest that the area initially designated as Point Maskelyne was a slightly elevated area. It is tempting to suggest that Dawes' observatory was built on its high point. However, we recommend due caution in interpreting these drawings, given the absence of such an elevated area in the side views. In addition, such a location cannot be reconciled with Dawes' (1788a) own description of the observatory's siting against the bedrock.

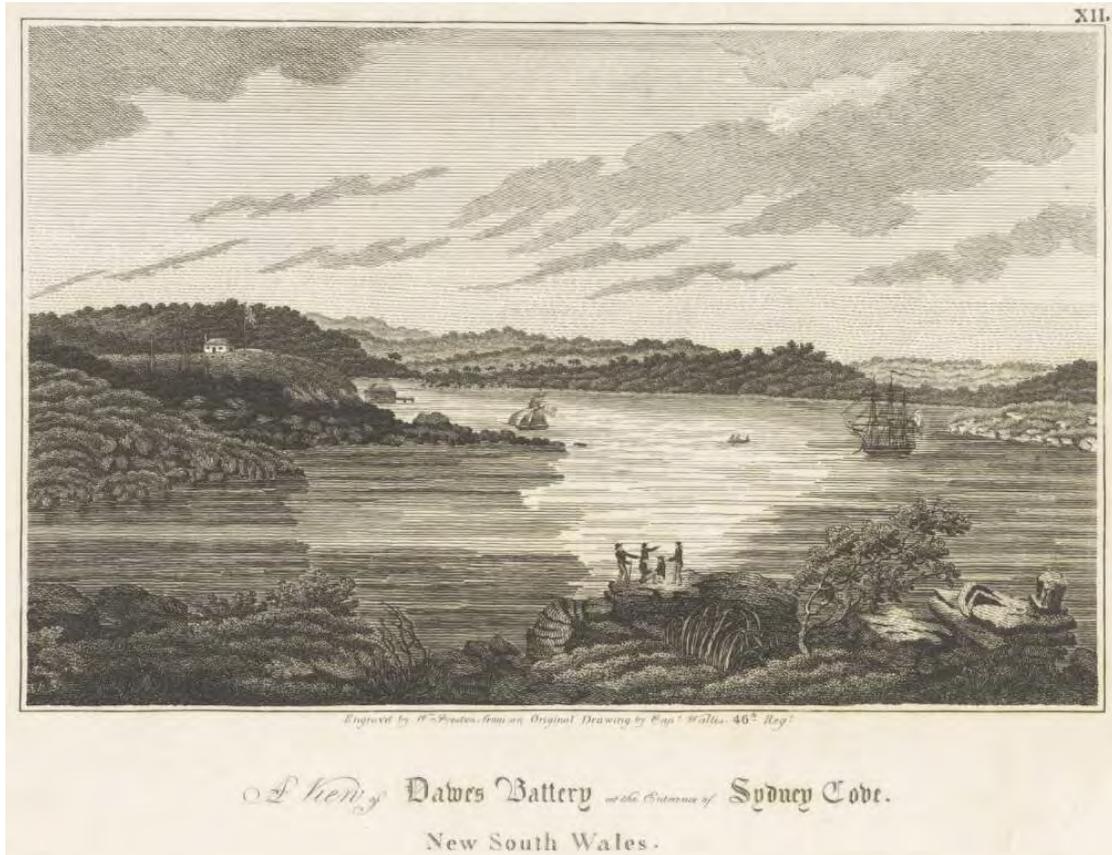

**Figure 22**: A view of Dawes Battery at the entrance of Sydney Cove, New South Wales, 1820 (National Library of Australia).

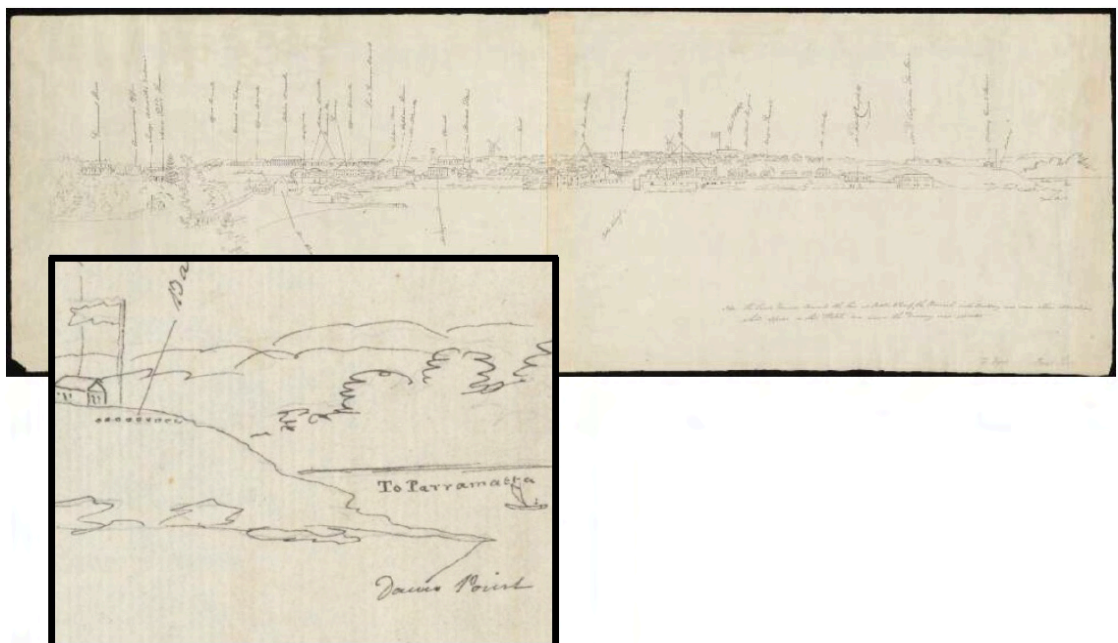

**Figure 23**: "Sydney Cove, east side, by John Eyre, ca. 1809." For a high-resolution image, see https://nla.gov.au/nla.obj-135179486/ (National Library of Australia).

The overall picture we have gained of Dawes' efforts in establishing an observatory in the new colony of New South Wales is one that is more intricate than most popular and many scholarly studies suggest. Dawes was not the first European astronomer to work his craft in the territory of New Holland. As we saw already, prior to his first tent-based observations, Lapérouse's astronomer, Dagelet, had already established the French expedition's portable observatory on the northern shores of Botany Bay. And, of course, Cook and Green had also carried out astronomical observations while at Botany Bay in 1770 (e.g., David, 1984). Even the concept of "Dawes' observatory", that is, the first British permanent astronomical building in the southern hemisphere, is fraught. As we have seen, after having established a timber and canvas structure consisting of two buildings by July 1788, there is evidence suggesting that the initial structure proved too confined and a second observatory, presumably made of stone, appears to have been under construction a mere year after observations began from the original building.

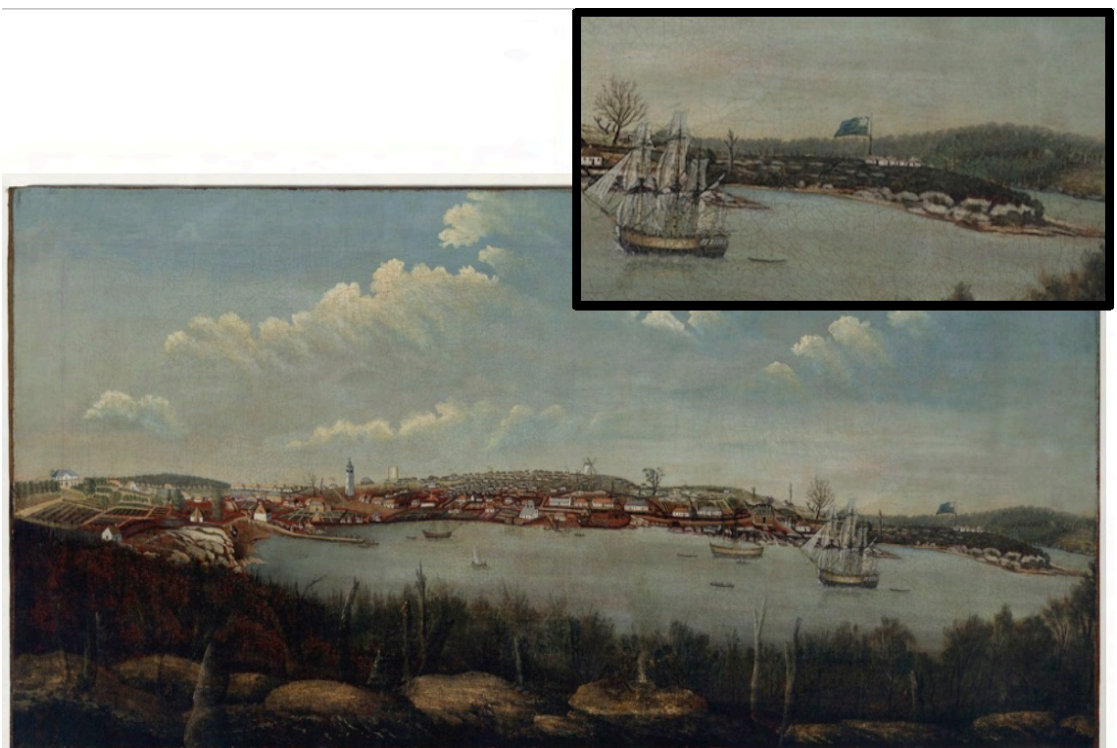

**Figure 24**: Sydney Cove looking to the west, ca. 1800; artist unknown (Mitchell Library, State Library of New South Wales).

Dawes was clearly an interesting person, highly intellectual and cultured. Of all issues he wrote about during his lifetime, he was least inclined to write about himself. His serious altercation with Governor Phillip halted what might have been a bright path forward as one of the fledgling colony's intellectual leaders. In any case, Dawes apparently buried his astronomical interests with his departure from the colony (although his son, William Rutter Dawes, developed into a formidable astronomer in his own right; e.g., Denning, 1913). He moved on in his career, finding new pursuits in working to abolish slavery in his role as Governor of Sierra Leone as well as in Antigua, West Indies, in later life. Yet, despite all of his achievements, William Dawes remains largely a 'mystery man'.

## 6 NOTES

[1] The term 'First Fleet' is a retrospective designation; at the time the fleet was made ready for departure, there were no plans for any subsequent fleets to be sent to the new colony of New South Wales (e.g., Hughes, 1987; Atkinson, 1997: 59–61).
[2] In the British Royal Navy's rating system at the time of the First Fleet's departure, a sixth-rate warship carried 20 to 28 carriage-mounted '9-pounder' guns on a single deck. Sixth-rate ships carrying fewer than 28 guns were categorised as 'post ships', so that they could still sail

under a full captain rather than a commander. The *Sirius* was an exception, in the sense that she only carried 10 guns yet was still classified as a sixth-rate vessel.

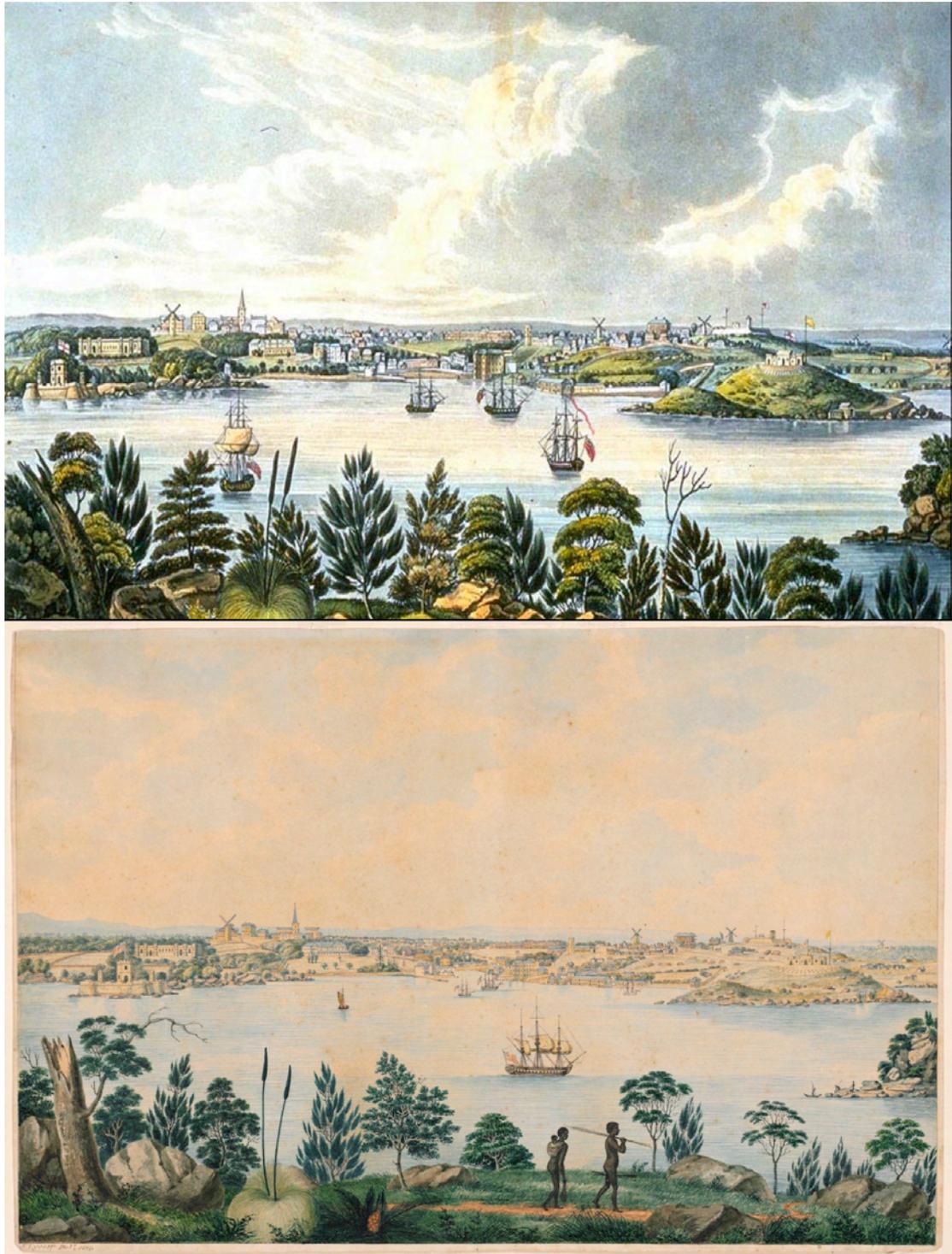

**Figure 25**: (*top*) "North view of Sidney, New South Wales", ca. 1825; Joseph Lycett (Mitchell Library, State Library of New South Wales). (*bottom*) "Two Forts and a Battery, 1822"; Joseph Lycett (Mitchell Library, State Library of New South Wales).

[3] Figure 1 is a portrait gallery of the main characters described in this article (except for William Dawes: see Figure 2), if and when their images were available in the public domain.
[4] In the late eighteenth century, the British Marine Corps were soldiers associated with the Royal Navy, not with military forces. They became *Royal* Marines in 1802.

[5] The British penal colony of New South Wales at the time of the First Fleet's voyage encompassed the area of present-day Australia east of the 135° East meridian, a North–South line bisecting present-day Australia about 20 km east of the South Australian town of Coober Pedy and ~115 km east of Alice Springs in the Northern Territory. Phillip's instructions of 25 April 1787 to proclaim the colony for Great Britain also included "all the islands adjacent in the Pacific Ocean" (King, 1998; Coltheart and the Museum of Australian Democracy, 2011) between latitude 10°37' South—Cape York in far north Queensland—and 43°39' South, the latitude of South East Cape, the southernmost point of the Tasmanian mainland. Incidentally, this area includes most of New Zealand.

[6] The adjective 'permanent' is appropriate in this context, for two reasons: (i) literature references to Dawes' observatory commonly identify the structure as 'permanent'; and (ii) although the actual building fell into disrepair only a few years after its construction, here the contrast of note is between tent observatories and facilities made of wood or bricks and mortar. The latter are longer-lived and hence 'permanent' is a more pertinent description.

[7] When Cook first arrived in the body of water now known as Botany Bay—called *Yarra* by the resident Bidjigal, Gweagal and Kameygal clans of the Eora Aboriginal language group (West, ca. 1882; Attenbrow, 2009)—he initially named it 'Stingrays Harbour': "The great quantity of these sort of fish found in this place occasioned my giving it the name of Stingrays Harbour" (Cook's log, 6 May 1770). Some time after his departure, however, he changed his mind: "The great quantity of plants [the expedition's botanists] Mr. [Joseph] Banks and Dr. [Daniel] Solander found in this place occasioned my giving it the name of ~~Botanist~~ Botany Bay" (Cook's correction; Beaglehole, 1968).

[8] Port Jackson encompasses Sydney Harbour, Middle Harbour, North Harbour and the Lane Cove and Parramatta Rivers. It was named by Cook as the H.M.B. *Endeavour* sailed by its entrance: "at noon we were ... about 2 or 3 miles from the land and abrest [*sic*] of a bay or harbour within there appeared to be a safe anchorage which I called Port Jackson." Cook named the harbour system after Sir George Jackson, Lord Commissioner of the British Admiralty and Judge Advocate of the Fleet.

[9] Dagelet's letter of 3 March 1788 (see below) contains the geographic coordinates of the French observatory (see also Barko, 2007; Morrison and Barko, 2009), which Dagelet obtained using the lunar distance method and de Lalande's astronomical tables (see, e.g., de Grijs, 2020). In doing so, he was assisted by De Roux d'Arbaud, his former student at the Royal Military Academy in Paris, whom he had selected in preference to another student, one Napoléon Bonaparte (see, e.g., Bartel, 1954: 133–134). In addition, King has recorded in his *Journal* that the expedition's priest, Claude-François Joseph Louis Receveur (1757–1788), "was buried near where the French had their observatory" (Morrison and Barko, 2009: 27–28, note 13). This gravesite is located a short walk east–northeastwards from the entrance of today's La Perouse Museum.

[10] Dagelet had been a member of the French Académie des Sciences since 1785 and a Professor of Mathematics at the Royal Military Academy in Paris since 1777.

[11] Although Dagelet's letter implies that it was delivered by an English seaman ("I profit from the presence of your seamen to send you my farewells"), there are no records of any English visit to Botany Bay at this time. Instead, in an excerpt from the logbook of the *Alexander*, we read "4 March 1788 – Several French officers came from Botany Bay" (Morrison and Barko, 2009: 27, note 7).

[12] Dawes (1788e) admitted to Maskelyne that when he was asked to take on the roles of engineer and artillery officer, he complained to Governor Phillip of the excessive workload compared with his Marine duties. Phillip was "very highly offended and several letters past [*sic*] between us". Although the disagreement dissolved of its own accord, it set the tone of the men's tense relationship during Dawes' tenure in New South Wales.

[13] One of Dawes' first tasks was to transfer the new settlement's ordnance from the *Sirius* to the shore, including the *Sirius*' guns. Wood (1924) explained, "... so eight of them were landed on the west point before the Observatory, and Lieutenant Dawes threw up a small breastwork in front of them, and afterwards built a platform for them" (see also Collins, 1798: 41, 166, 173 and 189).

[14] Steele (2005: 83) and Gibson (2012: 24) point out that 'dara' in the Eora language may simply be a suffix meaning 'here'. In his *Notebooks* on the Eora language, Dawes writes, "They were not speaking of Dara, for since, I have heard them repeat dara in the same word when I think they could not refer to that place. It seems to me to be peculiarly used when it is

spoken as of rowing <u>to a certain place to bring another back with you</u>. But this is mere conjecture" (Dawes' emphasis; Dawes, 1787–1788: *Notebook* A, 17).

[15] Dawes' carefully maintained meteorological journal was rediscovered among the Board of Longitude papers in 1977 (Morrison and Barko, 2009). His meticulous meteorological work set the standard for the implementation of meteorological duties at later state observatories, which provided a straightforward public justification for the associated costs (Haynes et al., 1996).

[16] The record accompanying this map in the collection of the Mitchell Library (State Library of New South Wales) states, "Hunter and Bradley made many charts and surveys during [their] eleven months' enforced stay on Norfolk Island [2 October 1788–8 May 1789]; this chart may have been amongst the number. … The data gathering for [this] chart may have taken place during five months of 1788, eight in 1789, and two in 1790" (https://search.sl.nsw.gov.au/permalink/f/lg5tom/SLNSW_ALMA21147000280002626).

[17] In 1926, the New South Wales Secretary (Minister) for Public Works and Minister for Railways was The Hon. Mr. Martin Matthew Flannery, M.L.A. (M.L.A.: Member of the Legislative Assembly of New South Wales).

[18] The distance of 90 feet indicated on the tablet already appeared in the *Daily Telegraph* of 29 May 1926.

**7 ACKNOWLEDGEMENTS**

We are indebted to the library staff at the Special Collections desk of the Mitchell Library, State Library of New South Wales (Sydney, Australia), the archivists at the State Archives of New South Wales (Kingswood, Australia) and the curator of The Rocks Museum. We are particularly grateful for support received from (in alphabetical order) Ben Arnsfield, Archivist, Data and Information Management, City of Sydney; Donna Newton, Librarian, Royal Australian Historical Society; Matthew Stuckings, Reference Librarian, Pictures and Manuscripts Section, National Library of Australia (Canberra); and Julie Sweeten (State Library of New South Wales, Special Collections Desk).